\documentstyle[11pt]{article}
\makeatletter \@addtoreset{equation}{section} \makeatother

\newcommand{\noi}{\vspace{12pt}\noindent}
\newcommand{\beq}{\begin{equation}}
\newcommand{\eeq}{\end{equation}}
\newcommand{\bea}{\begin{eqnarray}}
\newcommand{\eea}{\end{eqnarray}}

\newcommand{\e}[1]{{(\ref{#1})}}
\newcommand{\eq}[1]{{eq.\ (\ref{#1})}}
\newcommand{\es}[2]{{(\ref{#1}) and (\ref{#2})}}
\newcommand{\eqs}[2]{{eqs.\ (\ref{#1}) and (\ref{#2})}}
\newcommand{\Ref}[1]{{Ref.~\cite{#1}}}
\newcommand{\mb}[1]{{\mbox{${#1}$}}}
\newcommand{\equi}[1]{\stackrel{{#1}}{=}}

\newcommand{\ie}{{${ i.e.\ }$}}
\newcommand{\eg}{{${ e.g.\ }$}}

\newcommand{\cf}{{cf.\ }}
\newcommand{\wrt}{{with respect to }}
\newcommand{\wtho}{{with the help of }}

\newcommand{\aka}{{also known as }}

\newcommand{\bC}{\bar{C}}
\newcommand{\bF}{\bar{F}}
\newcommand{\bG}{\bar{G}}
\newcommand{\bj}{\bar{\jmath}}
\newcommand{\bell}{\bar{\ell}}
\newcommand{\bn}{\bar{n}}

\newcommand{\brho}{\bar{\rho}}
\newcommand{\bsigma}{\bar{\sigma}}
\newcommand{\bTheta}{\bar{\Theta}}
\newcommand{\x}{x}
\newcommand{\y}{y}
\newcommand{\z}{z}
\newcommand{\X}{X}
\newcommand{\Y}{Y}
\newcommand{\Z}{Z}
\newcommand{\eps}{\varepsilon^{}}
\newcommand{\D}{D}
\newcommand{\DD}{(D)}
\newcommand{\om}{\omega}
\newcommand{\Mext}{M_{\ext}^{}}
\newcommand{\ED}{E_{D}^{}}
\newcommand{\Eext}{E_{\ext}^{}}
\newcommand{\Wrho}{W_{\rho}^{}}
\newcommand{\Wbrho}{W_{\brho}^{}}
\newcommand{\WE}{W_{E}^{}}
\newcommand{\WED}{W_{\ED}^{}}
\newcommand{\WEext}{W_{\Eext}^{}}
\newcommand{\Xrho}{X_{\rho}^{}}
\newcommand{\Xbrho}{X_{\brho}^{}}

\newcommand{\XED}{X_{\ED}^{}}
\newcommand{\XEext}{X_{\Eext}^{}}
\newcommand{\Deltaone}{\Delta_{1}^{}}
\newcommand{\Deltarhoj}{\Delta_{\rho j}^{}}
\newcommand{\Deltarho}{\Delta_{\rho}^{}}
\newcommand{\Deltabrho}{\Delta_{\brho}^{}}
\newcommand{\DeltarhoE}{\Delta_{\rho,E}^{}}
\newcommand{\DeltarhoED}{\Delta_{\rho,\ED}^{}}
\newcommand{\DeltabrhoEext}{\Delta_{\brho,\Eext}^{}}
\newcommand{\DeltaE}{\Delta_{E}^{}}
\newcommand{\DeltaED}{\Delta_{\ED}^{}}
\newcommand{\DeltaEext}{\Delta_{\Eext}^{}}
\newcommand{\nurhoj}{\nu_{\rho j}^{}}
\newcommand{\nurho}{\nu_{\rho}^{}}
\newcommand{\nubrho}{\nu_{\brho}^{}}
\newcommand{\nurhoE}{\nu_{\rho,E}^{}}
\newcommand{\nurhoED}{\nu_{\rho,\ED}^{}}
\newcommand{\nubrhoEext}{\nu_{\brho,\Eext}^{}}

\newcommand{\ad}{{\rm ad}}

\newcommand{\sdet}{{\rm sdet}}

\newcommand{\ext}{{\rm ext}}

\newcommand{\Hf}{{1 \over 2}}

\newcommand{\Ih}{{i \over \hbar}}

\newcommand{\twotuborg}[2]{\left\{\begin{array}{c}{#1} \cr
                                {#2} \end{array} \right\}}
\newcommand{\threetuborg}[3]{\left\{\begin{array}{c}{#1} \cr
                             {#2} \cr {#3} \end{array} \right\}}
\newcommand{\deder}[1]{{ 
 {\stackrel{\raise.1ex\hbox{$\leftarrow$}}{\delta^r}   } 
\over {   \delta {#1}}  }}
\newcommand{\dedel}[1]{{ 
 {\stackrel{\lower.3ex \hbox{$\rightarrow$}}{\delta^l}   }
 \over {   \delta {#1}}  }}
\newcommand{\papar}[1]{{ 
 {\stackrel{\raise.1ex\hbox{$\leftarrow$}}{\partial^r}   } 
\over {   \partial {#1}}  }}
\newcommand{\papal}[1]{{ 
 {\stackrel{\lower.3ex \hbox{$\rightarrow$}}{\partial^l}   }
 \over {   \partial {#1}}  }}
\newcommand{\ddr}[1]{{ 
 {\stackrel{\raise.1ex\hbox{$\leftarrow$}}{\delta^r}   } 
\over {   \delta {#1}}  }}
\newcommand{\ddl}[1]{{ 
 {\stackrel{\lower.3ex \hbox{$\rightarrow$}}{\delta^l}   }
 \over {   \delta {#1}}  }}
\newcommand{\rpa}[1]{{ 
 \stackrel{\raise.1ex\hbox{$\leftarrow$}}{\partial^r_{#1}}   }}
\newcommand{\lpa}[1]{{ 
 \stackrel{\lower.3ex\hbox{$\rightarrow$}}{\partial^l_{#1}}  }}

\newcommand{\proofbox}{\begin{flushright}
${\,\lower0.9pt\vbox{\hrule \hbox{\vrule
height 0.2 cm \hskip 0.2 cm \vrule height 0.2 cm}\hrule}\,}$
\end{flushright}}

\newtheorem{theorem}{Theorem}[section]

\newtheorem{corollary}[theorem]{Corollary}
\newtheorem{definition}[theorem]{Definition}

\newtheorem{lemma}[theorem]{Lemma}

\newtheorem{proposition}[theorem]{Proposition}

\begin{document}
\thispagestyle{empty}
\title{\Large{\bf Semidensities, Second-Class Constraints\\ 
and Conversion in Anti-Poisson Geometry}}
\author{{\sc K.~Bering}$^1$\\Institute for Theoretical Physics \& Astrophysics
\\Masaryk University\\Kotl\'a\v{r}sk\'a 2\\CZ-611 37 Brno\\Czech Republic}
\maketitle
\vfill
\begin{abstract}
We consider Khudaverdian's geometric version of a Batalin-Vilkovisky (BV)
operator $\Delta_E$ in the case of a degenerate anti-Poisson manifold. The
characteristic feature of such an operator (aside from being a Grassmann-odd,
nilpotent, second-order differential operator) is that it sends semidensities
to semidensities. We find a local formula for the $\Delta_E$ operator in
arbitrary coordinates. As an important application of this setup, we consider
the Dirac antibracket on an antisymplectic manifold with antisymplectic
second-class constraints. We show that the entire Dirac construction,
including the corresponding Dirac BV operator $\Delta_{E_D}$, exactly follows
from conversion of the antisymplectic second-class constraints into
first-class constraints on an extended manifold.
\end{abstract}
\vfill
\begin{quote}
MCS number(s): 53A55; 58A50; 58C50; 81T70. 
\hrule width 5.cm \vskip 2.mm \noindent 
$^{1}${\small E-mail:~{\tt bering@physics.muni.cz}} \\ 
\end{quote}

\newpage

\tableofcontents

\section{Introduction}
\label{secintro}

\noi
Consider an antisymplectic manifold \mb{(M;E)} with coordinates
\mb{\Gamma^{A}}. Such structure was first used by Batalin and Vilkovisky to
quantize Lagrangian gauge theories \cite{bv81,bv83,bv84}. In general,
antisymplectic geometry has many of the characteristic features of ordinary
symplectic geometry, \eg the Jacobi identity and the Darboux Theorem, but there
are also important differences: There are no canonical volume form and no
Liouville Theorem in antisymplectic geometry \cite{b97}. In the covariant
Batalin-Vilkovisky (BV) formalism \cite{schwarz93,bt93} from around 1992 one is
(among other things) instructed to make separate choices of a measure density
\mb{\rho\!=\!\rho(\Gamma)} and a quantum action
\mb{\Wrho\!=\!\Wrho(\Gamma)}. However, the division into measure and
action part is to a large extent an arbitrary division, \ie it is always
possible to shift parts of the measure \mb{\rho} into the action
\mb{\Wrho} and vice versa. It is only a particular combination of these
two quantities, namely the Boltzmann semidensity
\beq
\exp[\Ih \WE]~\equiv~\sqrt{\rho}\exp[\Ih \Wrho] 
\label{boltzmannsemidensity}
\eeq
that enters the physical partition function \mb{{\cal Z}}. {}For instance, if 
there exist global Darboux coordinates
\mb{\Gamma^{A}\!=\!\{\phi^{\alpha};\phi^{*}_{\alpha}\}}, the partition function
reads
\beq
 {\cal Z}~=~\int \! [d\phi]\left.\exp[\Ih\WE] 
\right|_{\phi^{*}=\frac{\partial\psi}{\partial\phi}}~,\label{partitionfct}
\eeq
where \mb{\psi\!=\!\psi(\phi)} is the gauge fermion. (More generally, the
partition function \mb{{\cal Z}} is described by the so-called \mb{W}-\mb{X}
formalism \cite{bbd96,bbd06}.) The field-antifield formalism was reformulated
in \Ref{b06} entirely in the minimal language of semidensities, which skips
\mb{\rho} altogether. According to this minimal approach, the Boltzmann
semidensity \mb{\exp[\Ih\WE]} should satisfy the Quantum Master Equation
\beq
\DeltaE\exp[\Ih\WE]~=~0 \label{qmee}
\eeq
to ensure independence of gauge-fixing. Here \mb{\DeltaE} is Khudaverdian's BV
operator, which takes semidensities to semidensities, \cf
\Ref{bbd06,k99,kv02,k02,k04} and Definition~\ref{defkhudeltaesigma} below. 
Of course, the density \mb{\rho} may always be re-introduced to compare with
the 1992 formulation. In doing so, for an arbitrary choice of \mb{\rho}, 
\begin{enumerate}
\item
the Boltzmann semidensity \mb{\exp[\Ih \WE]} descend to a Boltzmann scalar
\mb{\exp[\Ih\Wrho]=\exp[\Ih\WE] / \sqrt{\rho}}, 
\item
the \mb{\DeltaE} operator descend to a (not necessarily nilpotent) odd
Laplacian \mb{\Deltarho}, which takes scalars to scalars, \cf 
Definition~\ref{defdeltarho} below; and
\item
the Quantum Master Eq.\ \e{qmee} descend to the Modified Quantum Master
Equation
\beq
(\Deltarho+\nurho)\exp[\Ih\Wrho]~=~0~,\label{mqme}
\eeq
where \mb{\nurho} is an odd scalar, \cf Definition~\ref{defnurho} below.
\end{enumerate}

\noi
We emphasize that this construction works for any \mb{\rho}. However, to arrive
at the 1992 formulation \cite{schwarz93,bt93}, which has \mb{\nurho\!=\!0} and
a nilpotent odd Laplacian \mb{\Delta_{\rho}^{2}\!=\!0}, one should impose
conditions on \mb{\rho}. 

\noi
The paper is organized as follows. Anti-Poisson geometry is reviewed in
Section~\ref{secantipoissongeom}. The notions of compatible two-form fields 
and bi-Darboux coordinates are introduced in Subsection~\ref{secantipoisson}. A
new Theorem~\ref{theorembidarboux} provides necessary and sufficient conditions
for the existence of bi-Darboux coordinates. The definition of the \mb{\DeltaE}
operator for a degenerate anti-Poisson structure \mb{E} is given using both
Darboux and general coordinates in Subsection~\ref{secsemi} and \ref{secarbi},
respectively. The \mb{\DeltaE} formula in general coordinates does require the
existence of a compatible two-form fields, however, it does not matter which
compatible two-form field that is used (in case there is more than one choice),
\cf Lemma~\ref{lemma2ourdeltaesigma}. All information about how the
\mb{\DeltaE} operator acts on semidensities can be packed into a Grassmann-odd
scalar quantity \mb{\nurho}, which already appeared in \eq{mqme} above. The
odd scalar \mb{\nurho} is important, because in practice it is easier to handle
a scalar object rather than the full second-order differential operator
\mb{\DeltaE}, and hence many of the ensuring arguments is performed using
\mb{\nurho}. The Dirac antibracket is an important application of the geometric
setup from Section~\ref{secantipoissongeom}, since it always admits a
compatible two-form field. Antisymplectic second-class constraints and the
Dirac antibracket \cite{bt93,bbd06,bbd97} are reviewed in
Subsection~\ref{secdirac}. A Proposition~\ref{propositionnurhod} in
Subsection~\ref{secdiracop} provides a useful formula for the corresponding
Dirac odd scalar \mb{\nurhoED}. Subsection~\ref{secreparam} discusses the
stability of the Dirac construction under reparameterizations of the
second-class constraints. In Section~\ref{secconv} the Dirac construction is
derived via conversion \cite{bf87,bff89,bt91,fl94,bm97} of the antisymplectic
second-class constraints into first-class constraints on an extended manifold.
As an application of the construction to Batalin-Vilkovisky quantization, the
corresponding Dirac and extended partition functions are provided in
Subsections~\ref{secpathintd} and \ref{secpathintext}, respectively. {}Finally,
Section~\ref{secconc} contains our conclusions.

\noi
{\em General remark about notation}. We have two types of grading: A Grassmann
grading \mb{\eps} and an exterior form degree \mb{p}. The sign conventions
are such that two exterior forms \mb{\xi} and \mb{\eta}, of Grassmann parity
\mb{\eps_{\xi}}, \mb{\eps_{\eta}} and exterior form degree
\mb{p_{\xi}}, \mb{p_{\eta}}, respectively, commute in the following graded
sense
\beq
 \eta \wedge \xi
~=~(-1)^{\eps_{\xi}\eps_{\eta}+p_{\xi}p_{\eta}}\xi\wedge\eta
\eeq
inside the exterior algebra.
We will often not write the exterior wedges ``\mb{\wedge}'' explicitly.

\section{Anti-Poisson Geometry}
\label{secantipoissongeom}

\subsection{Antibracket and Compatible Two-Form}
\label{secantipoisson}

\noi
We consider an anti-Poisson manifold \mb{(M;E^{AB})} with a (possibly
degenerate) antibracket
\beq
(F,G)~=~(F\rpa{A}) E^{AB}(\lpa{B}G)
~=~-(-1)^{(\eps_{F}+1)( \eps_{G}+1)}(G,F)~,~~~~~~~~
\lpa{A}~\equiv~\papal{\Gamma^{A}}~,
\label{antibracket}
\eeq
Here the \mb{\Gamma^{A}}'s denote local coordinates of Grassmann parity 
\mb{\eps_{A}\equiv\eps(\Gamma^{A})}, and \mb{E^{AB}\!=\!E^{AB}(\Gamma)}
is the local matrix representation of the anti-Poisson structure \mb{E}.
The Jacobi identity 
\beq
 \sum_{{\rm cycl.}~F,G,H}(-1)^{(\eps_{F}+1)( \eps_{H}+1)}
(F,(G,H)) ~=~0 \label{jacid}
\eeq
reads in local coordinates
\beq
\sum_{{\rm cycl.}~A,B,C}(-1)^{(\eps_{A}+1)( \eps_{C}+1)}
E^{AD} (\lpa{D} E^{BC}) ~=~0~. \label{ejacid}
\eeq
The main new feature (as compared to \Ref{b06}) is that the anti-Poisson
structure \mb{E^{AB}} could be degenerate. There is an anti-Poisson analogue of
Darboux's Theorem that states that locally, if the rank of \mb{E^{AB}} is
constant, there exist Darboux coordinates
\mb{\Gamma^{A}\!=\!\left\{\phi^{\alpha};\phi^{*}_{\alpha};\Theta^{a}\right\}},
such that the only non-vanishing antibrackets between the coordinates are
\mb{(\phi^{\alpha},\phi^{*}_{\beta})=\delta^{\alpha}_{\beta}
=-(\phi^{*}_{\beta},\phi^{\alpha})}. In other words, the Jacobi identity
is the integrability condition for the Darboux coordinates.
The variables \mb{\phi^{\alpha}}, \mb{\phi^{*}_{\alpha}} and \mb{\Theta^{a}}
are called {\em fields}, {\em antifields} and {\em Casimirs}, respectively.

\noi
We shall assume that the anti-Poisson manifold \mb{(M;E^{AB})} admits a
globally defined odd two-form field \mb{E_{AB}} with lower indices that is 
{\em compatible} with the anti-Poisson structure \mb{E^{AB}} in the sense that
\bea
 E^{AB}E_{BC}E^{CD}&=&E^{AD}~,\cr
 E_{AB}E^{BC}E_{CD}&=&E_{AD}~.\label{tripleeee}
\eea
As always, the matrices \mb{E^{AB}} and \mb{E_{AB}} are assumed to have the
Grassmann gradings
\beq
\eps(E^{AB})~=~\eps_{A}+ \eps_{B}+1~=~\eps(E_{AB})~,
\label{egrad}
\eeq
and the skew-symmetries
\bea
E^{BA}&=&-(-1)^{(\eps_{A}+1)(\eps_{B}+1)} E^{AB}~, \cr
E_{BA}&=&-(-1)^{\eps_{A}\eps_{B}}E_{AB}~.
\label{esym}
\eea
The odd two-form field can be written as
\beq
E~=~\Hf d\Gamma^{A}~E_{AB}~d\Gamma^{B}
~=~-\Hf E_{AB}~d\Gamma^{B}~d\Gamma^{A}~.
\eeq
The two-form field \mb{E_{AB}} would be closed if
\beq
 dE~=~0~, \label{eclosed}
\eeq
or equivalently, with all the indices written out, if
\beq
\sum_{{\rm cycl.}~A,B,C}(-1)^{\eps_{A} \eps_{C}}
(\lpa{A}E_{BC} ) ~=~0~. \label{eclosedabc} 
\eeq
A closed degenerate two-form is called a pre-antisymplectic structure. In the
non-degenerate case, the matrix \mb{E_{AB}} from \eq{tripleeee} would be a
closed antisymplectic two-form field and the inverse of the anti-Poisson
structure \mb{E^{AB}}. In the degenerate case, there is in general {\em not a
unique} matrix \mb{E_{AB}} fulfilling eqs.\ \e{tripleeee}, \es{egrad}{esym},
and there is {\em no} reason for it to be closed. In Darboux coordinates
\mb{\Gamma^{A}\!=\!\left\{\phi^{\alpha};\phi^{*}_{\alpha};\Theta^{a}\right\}}, 
there is still a freedom in a compatible two-form
\beq
E~=~d\phi^{*}_{\alpha}\wedge d\phi^{\alpha}
+d\Theta^{a} M_{a\alpha}\wedge d\phi^{\alpha}
+d\phi^{*}_{\alpha}N^{\alpha}{}_{a}\wedge d\Theta^{a}
+d\Theta^{a} M_{a\alpha} N^{\alpha}{}_{b}\wedge d\Theta^{b}
\label{emn}
\eeq
given by two arbitrary matrices \mb{M_{a\alpha}\!=\!M_{a\alpha}(\Gamma)} and
\mb{N^{\alpha}{}_{a}\!=\!N^{\alpha}{}_{a}(\Gamma)}. A Darboux coordinate system
 \mb{\Gamma^{A}\!=\!\left\{\phi^{\alpha};\phi^{*}_{\alpha};\Theta^{a}\right\}}
is called a {\em bi-Darboux} coordinate system,  if the two-form is just 
\mb{E=d\phi^{*}_{\alpha}\wedge d\phi^{\alpha}}, \ie if both the matrices
\mb{M_{a\alpha}\!=\!0} and \mb{N^{\alpha}{}_{a}\!=\!0} in \eq{emn} are equal to
zero. In short, the \mb{\Gamma^{A}}'s are bi-Darboux coordinates, if 
both matrices \mb{E^{AB}} and \mb{E_{AB}} with upper and lower indices are on 
standard form.

\begin{theorem}
Given an anti-Poisson manifold \mb{(M;E^{AB})} with a compatible two-form field
\mb{E_{AB}}. Then there locally exist bi-Darboux coordinates if and only if the
two-form field \mb{E_{AB}} is closed.
\label{theorembidarboux}
\end{theorem}

\noi
There is a similar Bi-Darboux Theorem for even Poisson structures. A proof
of Theorem~\ref{theorembidarboux} is given in Appendix~\ref{appbidarboux}.
One can define a projection as
\beq
P^{A}{}_{C}~\equiv~ E^{AB}E_{BC}~,
\eeq
or equivalently,
\beq
P_{A}{}^{C}~\equiv~ E_{AB}E^{BC} ~=~ (-1)^{\eps_{A}(\eps_{C}+1)}P^{C}{}_{A}~.
\eeq
It follows from property \e{tripleeee} that 
\beq
P^{A}{}_{B} P^{B}{}_{C}~=~P^{A}{}_{C}~.
\eeq
In the non-degenerate case \mb{P^{A}{}_{B}=\delta^{A}_{B}=P_{B}{}^{A}}.

\subsection{Odd Laplacian \mb{\Deltarho} on Scalars}
\label{secoddlapl}

\noi
Recall that a scalar function \mb{F\!=\!F(\Gamma)}, a density 
\mb{\rho\!=\!\rho(\Gamma)} and a semidensity \mb{\sigma\!=\!\sigma(\Gamma)} 
are by definition quantities that transform as
\beq
{}F~~\longrightarrow~F^{\prime}~=~F~,~~~~~~~~~~~
\rho~~\longrightarrow~~\rho^{\prime}~=~\frac{\rho}{J}~,~~~~~~~~~~~
\sigma~~\longrightarrow~~\sigma^{\prime}~=~\frac{\sigma}{\sqrt{J}}~,
\label{coordtransf}
\eeq
respectively, under general coordinate transformations
\mb{\Gamma^{A}\to\Gamma^{\prime A}}, where 
\mb{J\equiv\sdet\frac{\partial \Gamma^{\prime A}}{\partial \Gamma^{B}}}
denotes the Jacobian.  We shall ignore the global issues of orientation
and choice of square root. Also we assume that densities \mb{\rho} are
invertible.

\begin{definition}
Given a choice of a density \mb{\rho}, the {\bf odd Laplacian} \mb{\Deltarho} 
is defined as \cite{bt93}
\beq
\Deltarho~\equiv~\frac{(-1)^{\eps_{A}}}{2\rho}
\lpa{A}\rho E^{AB}\lpa{B}~.\label{deltarho}
\eeq
\label{defdeltarho}
\end{definition}

\noi
This Grassmann-odd, second-order operator takes scalar functions to scalar
functions. In situations with more than one anti-Poisson structure \mb{E^{AB}},
we shall sometimes use the slightly longer notation
\mb{\Deltarho\equiv\DeltarhoE} to acknowledge that it depends on two inputs: 
\mb{\rho} and \mb{E^{AB}}. The odd Laplacian \mb{\Deltarho} ``differentiates''
the antibracket \mb{(\cdot,\cdot)}, \ie the following Leibniz-type rule holds
\beq
\Deltarho(F,G)~=~(\Deltarho F,G)
+(-1)^{(\eps_{F}+1)}(F,\Deltarho G)~. \label{deltaleibnizantib}
\eeq
{}For further information on this important operator, see \Ref{bbd06,b06} and 
Subsection~\ref{secnilpdeltarho} below.

\subsection{The \mb{\DeltaE} Operator on Semidensities}
\label{secsemi}

\noi
There is an another important Grassmann-odd, nilpotent, second-order operator 
\mb{\DeltaE} that depends only on the anti-Poisson structure \mb{E^{AB}}.
Contrary to the odd Laplacian \mb{\Deltarho\equiv\DeltarhoE} of last
Subsection~\ref{secoddlapl}, the \mb{\DeltaE} operator does {\em not} rely on a
choice of density \mb{\rho}. The caveat is that while the odd Laplacian
\mb{\Deltarho} takes scalars to scalars, the \mb{\DeltaE} operator takes
semidensities to semidensities of opposite Grassmann parity. Equivalently, the
\mb{\DeltaE} operator transforms as
\beq
\DeltaE~~\longrightarrow~\Delta_{E}^{\prime}
~=~\frac{1}{\sqrt{J}}\DeltaE\sqrt{J} 
\label{deltaetransf}
\eeq
under general coordinate transformations \mb{\Gamma^{A}\to\Gamma^{\prime A}},
cf.\ \eq{coordtransf}. It is defined as follows:

\begin{definition}
Let there be given an anti-Poisson manifold \mb{(M;E)}.
In Darboux coordinates \mb{\Gamma^{A}}, the \mb{\DeltaE} {\bf operator} is
defined on a semidensity \mb{\sigma} as \cite{bbd06,k99,kv02,k02,k04}
\beq
(\DeltaE\sigma)~\equiv~(\Deltaone\sigma)~,\label{khudeltaesigma}
\eeq
where \mb{\Deltaone} denotes the expression \e{deltarho} for the odd
Laplacian \mb{\Delta_{\rho=1}^{}} with \mb{\rho} replaced by \mb{1}.
\label{defkhudeltaesigma}
\end{definition}

\noi
It is implicitly understood in \eq{khudeltaesigma} that the formula for
the \mb{\Deltaone} operator \e{deltarho} and the semidensity \mb{\sigma}
both refer to the same Darboux coordinates \mb{\Gamma^{A}}. The parentheses in
\eq{khudeltaesigma} indicate that the equation should be understood as an
equality among semidensities (in the sense of zeroth-order differential
operators) rather than an identity among differential operators. The
Definition~\ref{defkhudeltaesigma} does not depend on the Darboux coordinate
system being used, due to the following Lemma~\ref{lemmakhudeltaesigma}:

\begin{lemma}
When using the Definition~\ref{defkhudeltaesigma}, the 
\mb{(\DeltaE\sigma)} transforms as a semidensity under (anticanonical)
transformations between sets of Darboux coordinates.
\label{lemmakhudeltaesigma}
\end{lemma}

\noi
Thus the \mb{\DeltaE} operator is a well-defined operator on an open cover of
Darboux neighborhoods. Within this cover, the \mb{\DeltaE} is indirectly
defined in non-Darboux coordinates by use of the transformation property
\e{deltaetransf}. Lemma~\ref{lemmakhudeltaesigma} was first proven in the
non-degenerate case in \Ref{k04} and in the degenerate case in \Ref{bbd06}. We
shall also give an independent proof in the next Subsection~\ref{secarbi}, \cf 
Lemma~\ref{lemmaourdeltaesigma} below. In some cases the \mb{\DeltaE} operator
may be extended to singular points (\ie points where the rank of the
anti-Poisson tensor \mb{E^{AB}} jumps) by continuity.

\noi
Working in Darboux coordinates, it is obvious that the \mb{\DeltaE} operator
super-commutes with itself, because the \mb{\Gamma^{A}}-derivatives have no
\mb{\Gamma^{A}}'s to act on when \mb{E^{AB}} is on Darboux form. Therefore 
\mb{\DeltaE} is nilpotent,
\beq
\Delta_{E}^{2}~=~\Hf [\DeltaE,\DeltaE]~=~0~.\label{deltaenilp}
\eeq
Same sort of reasoning shows that \mb{\DeltaE=\Delta_{E}^{T}} is symmetric.

\subsection{The \mb{\DeltaE} Operator in General Coordinates}
\label{secarbi}

\noi
We now give a definition of the \mb{\DeltaE} operator that does not refer
to Darboux coordinates. 

\begin{definition}
Given an anti-Poisson manifold \mb{(M;E^{AB})} that admits a compatible 
two-form field \mb{E_{AB}}. In arbitrary coordinates \mb{\Gamma^{A}}, the
\mb{\DeltaE} operator is defined as
\beq
 (\DeltaE\sigma)~\equiv~ (\Deltaone\sigma)
+\left(\frac{\nu^{(1)}}{8}-\frac{\nu^{(2)}}{8}-\frac{\nu^{(3)}}{24}
+\frac{\nu^{(4)}}{24}+\frac{\nu^{(5)}}{12}\right)\sigma~,
\label{ourdeltaesigma}
\eeq
where
\bea
\nu^{(1)}&\equiv&
(-1)^{\eps_{A}}(\lpa{B}\lpa{A}E^{AB})~,\label{nu1} \\
\nu^{(2)}&\equiv& (-1)^{\eps_{A}\eps_{C}}(\lpa{D}E^{AB})E_{BC}
(\lpa{A}E^{CD})~,\label{nu2} \\
\nu^{(3)}&\equiv&(-1)^{\eps_{B}}(\lpa{A}E_{BC})
E^{CD}(\lpa{D}E^{BA})~,\label{nu3} \\
\nu^{(4)}&\equiv&(-1)^{\eps_{B}}(\lpa{A}E_{BC})
E^{CD}(\lpa{D}E^{BF})P_{F}{}^{A}~,\label{nu4} \\
\nu^{(5)}&\equiv&(-1)^{\eps_{A}\eps_{C}}(\lpa{D}E^{AB})E_{BC}
(\lpa{A}E^{CF})P_{F}{}^{D} \cr
&=&(-1)^{(\eps_{A}+1)\eps_{B}} E^{AD}(\lpa{D}
E^{BC})(\lpa{C}E_{AF})P^{F}{}_{B}~.\label{nu5} 
\eea
\label{defourdeltaesigma}
\end{definition}

\noi 
Notice that in Darboux coordinates, where \mb{E^{AB}} is constant, \ie
independent of the coordinates \mb{\Gamma^{A}}, the last five terms
\mb{\nu^{(1)}}, \mb{\nu^{(2)}}, \mb{\nu^{(3)}}, \mb{\nu^{(4)}} and
\mb{\nu^{(5)}} become zero. Hence the new Definition~\ref{defourdeltaesigma}
agrees in Darboux coordinates with the previous
Definition~\ref{defkhudeltaesigma}.
The benefit of the new Definition~\ref{defourdeltaesigma} is that one now have 
an explicit formula for \mb{\DeltaE} in an arbitrary coordinate system.
The full justification of Definition~\ref{defourdeltaesigma} is provided by 
the following Lemma~\ref{lemmaourdeltaesigma} and 
Lemma~\ref{lemma2ourdeltaesigma}.

\begin{lemma} When using the new Definition~\ref{defourdeltaesigma}, the 
\mb{(\DeltaE\sigma)} transforms as a semidensity under general coordinate
transformations.
\label{lemmaourdeltaesigma}
\end{lemma}

\begin{lemma} When using the new Definition~\ref{defourdeltaesigma}, the 
\mb{(\DeltaE\sigma)} does not depend on the compatible two-form field
\mb{E_{AB}} used.
\label{lemma2ourdeltaesigma}
\end{lemma}

\noi
The explicit formula \e{ourdeltaesigma} and Lemma~\ref{lemmaourdeltaesigma} 
are the main results of Section~\ref{secantipoissongeom}. 

\noi
{\sc Proof of Lemma~\ref{lemma2ourdeltaesigma}}:~~The two-form field
\mb{E_{AB}} enters only the Definition~\ref{defourdeltaesigma} via
\mb{\nu^{(2)}}, \mb{\nu^{(3)}}, \mb{\nu^{(4)}} and \mb{\nu^{(5)}}. Assuming the
Lemma~\ref{lemmaourdeltaesigma}, \ie that the behavior \e{deltaetransf} under
general coordinate transformations has already been established, one may, in
particular, go to Darboux coordinates, where \mb{\nu^{(2)}}, \mb{\nu^{(3)}},
\mb{\nu^{(4)}} and \mb{\nu^{(5)}} vanish identically.
\proofbox

\noi
To prove Lemma~\ref{lemmaourdeltaesigma} we shall first reformulate it as an
equivalent Lemma~\ref{lemmanurho}, \cf below. We shall also only explicitly
consider the case where \mb{\sigma} is invertible to simplify the presentation.
(The non-invertible case is fundamentally no different.) In the invertible
case, we customarily write the semidensity \mb{\sigma=\sqrt{\rho}} as a square
root of a density \mb{\rho}, and define a Grassmann-odd quantity
\mb{\nurho} as follows.

\begin{definition}
The {\bf odd scalar} \mb{\nurho} is defined as
\beq
\nurho~\equiv~ \frac{1}{\sqrt{\rho}}(\DeltaE\sqrt{\rho})
~=~ \nu_{\rho}^{(0)}+\frac{\nu^{(1)}}{8}-\frac{\nu^{(2)}}{8}
-\frac{\nu^{(3)}}{24}+\frac{\nu^{(4)}}{24}+\frac{\nu^{(5)}}{12}~,
\label{nurho}
\eeq
where \mb{\nu^{(1)}}, \mb{\nu^{(2)}}, \mb{\nu^{(3)}}, \mb{\nu^{(4)}},
\mb{\nu^{(5)}} are given in eqs.\ \e{nu1}--\e{nu5}, and the quantity
\mb{\nu_{\rho}^{(0)}} is given as
\beq
\nu_{\rho}^{(0)}~\equiv~\frac{1}{\sqrt{\rho}}(\Deltaone\sqrt{\rho})~.
\label{nurho0}
\eeq
\label{defnurho}
\end{definition}

\noi
In situations with more than one anti-Poisson structure \mb{E^{AB}}, we shall
sometimes use the slightly longer notation \mb{\nurho\equiv\nurhoE}. By
dividing both sides of the definition \e{ourdeltaesigma} with the semidensity
\mb{\sigma}, one may reformulate the content of
Lemma~\ref{lemmaourdeltaesigma} as:

\begin{lemma}
The Grassmann-odd quantity \mb{\nurho} is a scalar, \ie it does not depend on 
the coordinate system.
\label{lemmanurho}
\end{lemma}

\noi
We shall give two independent proofs of this important Lemma~\ref{lemmanurho};
one relying on Darboux Theorem and the other using infinitesimal coordinate
transformations. 

\noi
{\sc Proof of Lemma~\ref{lemmanurho} using a Darboux coordinate patch}:~~
It is enough to consider how \mb{\nurho} behaves on coordinate transformations
\mb{\Gamma^{A}_{0}\to\Gamma^{A}} between Darboux coordinates
\mb{\Gamma^{A}_{0}} and general coordinates \mb{\Gamma^{A}}. (An arbitrary
coordinate transformation between two general coordinate patches can always be
split into two successive coordinate transformations of the above kind by
inserting a third Darboux coordinate patch in between.) The idea is now to
first consider the expression \e{nurho} for \mb{\nurho} in the \mb{\Gamma^{A}}
coordinate system, and decompose it in building blocks that refer to the
Darboux coordinates \mb{\Gamma^{A}_{0}}, \eg
\beq
E^{AD}~=~(\Gamma^{A} \papar{\Gamma^{B}_{0}}) 
E^{BC}_{~0}(\papal{\Gamma^{C}_{0}}\Gamma^{D})~,~~~~~
E_{AD}~=~(\papal{\Gamma^{a}}\Gamma^{B}_{0})
E_{BC}^{~0}(\Gamma^{C}_{0}\papar{\Gamma^{D}})~,~~~~~
\rho~=~\frac{\rho_{0}^{}}{J}~.
\eeq
Here \mb{J\equiv\sdet(\partial \Gamma^{A} / \partial \Gamma^{B}_{0})} denotes
the Jacobian of the coordinate transformations
\mb{\Gamma^{A}_{0}\to\Gamma^{A}}.  Recall that the two-form field
\mb{E_{BC}^{~0}} is not necessarily constant in the Darboux coordinates
\mb{\Gamma^{A}_{0}}, cf \eq{emn}. By straightforward calculation, one gets
\bea
\nu^{(0)}_{\rho}&\equiv&\frac{1}{\sqrt{\rho}}(\Delta_{1,E}^{}\sqrt{\rho})
~=~\frac{1}{\sqrt{\rho}}(\Delta_{J,E_{0}^{}}^{}\sqrt{\rho})
~=~\frac{1}{\sqrt{\rho_{0}^{}}}(\Delta_{1,E_{0}^{}}\sqrt{\rho_{0}^{}})
-\frac{1}{\sqrt{J}}(\Delta_{1,E_{0}^{}}\sqrt{J})~, \label{nurho0transfrule}\\
\nu^{(1)}&=&\frac{8}{\sqrt{J}}(\Delta_{1,E_{0}^{}}\sqrt{J})
-(-1)^{\eps_{B}}(\papal{\Gamma^{A}_{0}}\Gamma^{B},~
\papal{\Gamma^{B}}\Gamma^{A}_{0})~,\label{nu1transfrule}   \\
\nu^{(2)}&=&-(-1)^{\eps_{B}}(\papal{\Gamma^{A}_{0}}\Gamma^{B},~
\papal{\Gamma^{B}}\Gamma^{A}_{0})
-2(-1)^{\eps_{B}}(\papal{\Gamma^{A}_{0}}\Gamma^{B},~
\papal{\Gamma^{B}}\Gamma^{C}_{0})P_{C}^{0,A}~, \label{nu2transfrule}  \\
\nu^{(3)}&=&3(-1)^{\eps_{B}}(\papal{\Gamma^{A}_{0}}\Gamma^{B},~
\papal{\Gamma^{B}}\Gamma^{C}_{0})P_{C}^{0,A} \cr
&&-(-1)^{(\eps_{A}+1)(\eps_{C}+1)}(\Gamma^{A}_{0}\papar{\Gamma^{B}})
(\Gamma^{B}\papar{\Gamma^{C}_{0}},~E_{AD}^{~0})E^{DC}_{~0}
~, \label{nu3transfrule}  \\
\nu^{(4)}&=&(-1)^{\eps_{B}}(\papal{\Gamma^{A}_{0}}\Gamma^{B},~
\papal{\Gamma^{B}}\Gamma^{C}_{0})P_{C}^{0,A}+
2(-1)^{\eps_{A}\eps_{C}}P_{A}^{0,B}(\papal{\Gamma^{B}_{0}}\Gamma^{C},~
\Gamma^{A}_{0}\papar{\Gamma^{D}})P^{D}{}_{C}\cr
&&-(-1)^{(\eps_{A}+1)(\eps_{C}+1)}(\Gamma^{A}_{0}\papar{\Gamma^{B}})
(\Gamma^{B}\papar{\Gamma^{C}_{0}},~E_{AD}^{~0})E^{DC}_{~0}
~, \label{nu4transfrule}  \\
\nu^{(5)}&=&-2(-1)^{\eps_{B}}(\papal{\Gamma^{A}_{0}}\Gamma^{B},~
\papal{\Gamma^{B}}\Gamma^{C}_{0})P_{C}^{0,A}
-(-1)^{\eps_{A}\eps_{C}}P_{A}^{0,B}(\papal{\Gamma^{B}_{0}}\Gamma^{C},~
\Gamma^{A}_{0}\papar{\Gamma^{D}})P^{D}{}_{C}~. \label{nu5transfrule}
\eea
The last equality in \eq{nurho0transfrule} is a non-trivial property of the 
odd Laplacian. It is now easy to check that all but one of the above terms 
on the right-hand sides of eqs.\ \e{nurho0transfrule}--\e{nu5transfrule} cancel
in the pertinent linear combination \e{nurho}, \ie
\beq
 \nurho~=~ \nu_{\rho}^{(0)}+\frac{\nu^{(1)}}{8}-\frac{\nu^{(2)}}{8}
-\frac{\nu^{(3)}}{24}+\frac{\nu^{(4)}}{24}+\frac{\nu^{(5)}}{12}~=~
\frac{1}{\sqrt{\rho_{0}^{}}}(\Delta_{1,E_{0}^{}}\sqrt{\rho_{0}^{}})~.
\eeq
The surviving term, on the other hand, is just the definition for \mb{\nurho}
in the Darboux coordinates \mb{\Gamma^{A}_{0}}.
\proofbox

\noi
{\sc Proof of Lemma~\ref{lemmanurho} using infinitesimal coordinate
transformations}:~~Under an arbitrary infinitesimal coordinate transformation
\mb{\delta\Gamma^{A}=X^{A}}, one calculates
\bea
\delta\nu_{\rho}^{(0)}&=&-\Hf \Deltaone{\rm div}_{1}^{}X~,\label{dnurho0}\\
\delta\nu^{(1)}&=& 4 \Deltaone{\rm div}_{1}^{}X 
+ (-1)^{\eps_{A}}(\lpa{C}E^{AB})
(\lpa{B}\lpa{A}X^{C})~, \label{dnu1} \\
\delta\nu^{(2)}&=&(-1)^{\eps_{A}}(\lpa{D}E^{AB})
\left(2P_{B}{}^{C}(\lpa{C}\lpa{A}X^{D})+
(\lpa{B}\lpa{A}X^{C})P_{C}{}^{D}\right)~, \label{dnu2}\\
\delta\nu^{(3)}&=&(-1)^{\eps_{B}}(\lpa{A}E_{BC})
E^{CD}\left( (\lpa{D}X^{B}\rpa{F})E^{FA}
-(-1)^{(\eps_{A}+1)(\eps_{B}+1)}(\lpa{D}X^{A}
\rpa{F})E^{FB}\right) \cr
&&-\frac{3}{2}(-1)^{\eps_{A}}P_{C}{}^{D}(\lpa{D}E^{AB})
(\lpa{B}\lpa{A}X^{C})~, \label{dnu3} \\
\delta\nu^{(4)}&=&-2(-1)^{\eps_{B}}(\lpa{A}
\lpa{B}X^{C})P_{C}{}^{D}(\lpa{D}E^{BF})P_{F}{}^{A}\cr
&&+(-1)^{\eps_{B}}(\lpa{A}E_{BC})
E^{CD}(\lpa{D}X^{B}\rpa{F})E^{FA} \cr
&&+(-1)^{(\eps_{B}+1)\eps_{F}}P_{F}{}^{A}(\lpa{A}
E_{BC})E^{CD}(\lpa{D}X^{F}\rpa{G})E^{GB}\cr
&&+\Hf(-1)^{\eps_{A}}P_{C}{}^{D}(\lpa{D}
E^{AB})(\lpa{B}\lpa{A}X^{C})~, \label{dnu4} \\
\delta\nu^{(5)}&=&-(-1)^{\eps_{A}(\eps_{B}+1)}(\lpa{A}
E_{BC})E^{CD}(\lpa{D}X^{A}\rpa{F})E^{FB} \cr
&&+2(-1)^{\eps_{B}}(\lpa{A}\lpa{B}X^{C})P_{C}{}^{D}
(\lpa{D}E^{BF})P_{F}{}^{A}~. \label{dnu5}
\eea
A proof of \eqs{dnurho0}{dnu1} can be found in \Ref{b06}, and eqs.\ 
\e{dnu2}--\e{dnu5} are proven in Appendix~\ref{appnurho}.
One may verify that while the six constituents \mb{\nu_{\rho}^{(0)}},
\mb{\nu^{(1)}}, \mb{\nu^{(2)}}, \mb{\nu^{(3)}}, \mb{\nu^{(4)}} and 
\mb{\nu^{(5)}} separately have non-trivial transformation properties, the
linear combination \mb{\nurho} in \eq{nurho} is indeed a scalar.
\proofbox

\noi
The new Definition~\ref{defourdeltaesigma} is clearly symmetric 
\mb{\DeltaE=\Delta_{E}^{T}}. To check explicitly in general coordinates that
\mb{\DeltaE} is nilpotent is a straightforward (but admittedly tedious)
exercise. However, since we have just proven that \mb{\DeltaE} behaves
covariantly under general coordinate transformations, our previous proof of
nilpotency from last Subsection~\ref{secsemi} using Darboux coordinates
suffices. To summarize:

\begin{theorem}
The \mb{\DeltaE} operator \e{ourdeltaesigma} is nilpotent \e{deltaenilp} if and
only if the antibracket \e{antibracket} satisfies the Jacobi identity
\e{ejacid}.
\label{theoremnilpjac}
\end{theorem}

\noi
In the rest of the paper we will always assume that the Jacobi identity
\e{ejacid} is satisfied, and hence that the \mb{\DeltaE} operator 
\e{ourdeltaesigma} is nilpotent.

\subsection{Nilpotency Condition for the odd Laplacian \mb{\Deltarho}}
\label{secnilpdeltarho}

\noi
At this point it is instructive to recall the nilpotency condition for the
odd Laplacian \mb{\Deltarho}, although we shall not assume that it is 
satisfied. It follows from the Jacobi identity \e{ejacid} alone, that
\mb{\Delta_{\rho}^{2}} is a linear derivation, \ie a first-order differential
operator. The interplay between the two second-order differential operators
\mb{\DeltaE} and \mb{\Deltarho} is perhaps best summarized by the
following operator identity:
\beq
\Deltarho+\nurho~=~\frac{1}{\sqrt{\rho}}\DeltaE\sqrt{\rho}~,
\label{deltarhodeltaeopid}
\eeq
\cf eq.\ (5.9) of \Ref{b06}. In words: Apart from the \mb{\nurho} term the
odd Laplacian \mb{\Deltarho} is the \mb{\DeltaE} operator dressed with a
\mb{\sqrt{\rho}} factor. {}From this operator identity \e{deltarhodeltaeopid}
and the nilpotency \e{deltaenilp} of the \mb{\DeltaE} operator, one derives
the explicit form of the linear derivation:
\beq
 \Delta_{\rho}^{2}~=~(\nurho,~\cdot~)~.
\label{deltarhosquared}
\eeq
Therefore the nilpotency condition for \mb{\Deltarho} reads \cite{bbd06,kv02}
\beq
\Delta_{\rho}^{2}~=~0~~~~~~~~\Leftrightarrow~~~~~~~~
\nurho~{\rm is~a~Casimir.}
\eeq
Let us also mention for later that if one acts with the operator identity 
\e{deltarhodeltaeopid} on a scalar function \mb{\sqrt{F}}, one gets
\beq
\nu_{\rho F}^{}~=~\nurho+\frac{1}{\sqrt{F}}(\Deltarho\sqrt{F})~.
\label{nurhof}
\eeq

\subsection{Alternative Expressions}
\label{secalter}

\noi
It is convenient to introduce
\bea
\nu^{(23)}~\equiv&\nu^{(2)}+\nu^{(3)}
&=~(-1)^{\eps_{B}}(\lpa{A}
P_{B}{}^{C})(\lpa{C}E^{BA})~, \\
\nu^{(35)}~\equiv&\nu^{(3)}+\nu^{(5)}
&=~(-1)^{\eps_{B}}(\lpa{A}
P_{B}{}^{C})P_{C}{}^{D}(\lpa{D}E^{BA}) \cr
&&=~(-1)^{\eps_{B}}(\lpa{A}P^{B}{}_{C})E^{CD}
(\lpa{D}P_{B}{}^{A})~, \\
\nu^{(45)}~\equiv&\nu^{(4)}+\nu^{(5)}
&=~(-1)^{\eps_{B}}P_{A}{}^{D}
(\lpa{D}P_{B}{}^{C})(\lpa{C}E^{BA})~, \\\
\nu^{(23)}_{(45)}~\equiv&\nu^{(23)}-\nu^{(45)}
&=~(-1)^{\eps_{B}(\eps_{D}+1)}(\lpa{A}
P^{B}{}_{C})E^{CD}(\lpa{B}P_{D}{}^{A})~.
\eea
Then the \mb{\DeltaE} operator \e{ourdeltaesigma} may be re-written as
\bea
 (\DeltaE\sigma)&=&(\Deltaone\sigma)+\left(\frac{\nu^{(1)}}{8}
-\frac{\nu^{(2)}}{24}-\frac{\nu^{(23)}}{12}
+\frac{\nu^{(35)}+\nu^{(45)}}{24}\right)\sigma~,
\label{ourdeltaesigmaalter} \\
&=&(\Deltaone\sigma)+\left(\frac{\nu^{(1)}}{8}
-\frac{\nu^{(2)}+\nu^{(23)}-\nu^{(35)}+\nu^{(23)}_{(45)}}{24}\right)\sigma~.
\label{ourdeltaesigmaalteralter}
\eea
In the closed case \e{eclosed} one may show that
\beq
\nu^{(35)}+\nu^{(45)}~=~0~, 
\eeq
so that the \mb{\DeltaE} operator \e{ourdeltaesigmaalter} simplifies to
\beq
 (\DeltaE\sigma)~=~(\Deltaone\sigma)+\left(\frac{\nu^{(1)}}{8}
-\frac{\nu^{(2)}}{8}-\frac{\nu^{(3)}}{12}\right)\sigma~.
\label{ourdeltaesigmaclosed}
\eeq
In the non-degenerate case, which is automatically closed, one also has
\beq
 \nu^{(23)}~=~0~,
\eeq
so that the \mb{\DeltaE} operator \e{ourdeltaesigmaalter} simplifies
even further to
\beq
 (\DeltaE\sigma)~=~(\Deltaone\sigma)+\left(\frac{\nu^{(1)}}{8}
-\frac{\nu^{(2)}}{24}\right)\sigma~,
\label{ourdeltaesigmanondeg}
\eeq
in agreement with eq.\ (5.1) in \Ref{b06}.

\section{Second-Class Constraints}
\label{secsecondclass}

\subsection{Review of Dirac Antibracket}
\label{secdirac}

\noi
One of the most important examples of degenerate anti-Poisson structures is
provided by the Dirac antibracket \cite{bt93,bbd06,bbd97}. Consider a manifold
\mb{(M;E)} with a non-degenerate anti-Poisson structure \mb{E^{AB}} (called an 
antisymplectic phase space), and let a submanifold 
\mb{\tilde{M}\equiv\{\Gamma\in M|\Theta(\Gamma)=0\}} be the zero-locus of a
set of constraints \mb{\Theta^{a}\!=\!\Theta^{a}(\Gamma)} with Grassmann parity
\mb{\eps({\Theta^a})\!=\!\eps_{a}}. (In this Subsection, the defining set of
constraints is kept fixed for simplicity. We will consider reparametrizations 
of the constraints in Subsection~\ref{secreparam}.) Assume that the
\mb{\Theta^{a}} constraints are second-class in the antibracket sense, \ie
the antibracket matrix
\beq
E^{ab} ~\equiv~ (\Theta^{a},\Theta^{b}) \label{definingthetatheta}
\eeq
of the \mb{\Theta^{a}} constraints has by definition an inverse matrix 
\mb{E_{ab}},
\beq
E_{ab}E^{bc} ~=~ \delta^{c}_{a}~.
\eeq
The Dirac antibracket is defined completely analogous to the usual Dirac 
bracket for even Poisson brackets \cite{bt93}, 
\beq
(F,G)_{\D}^{} ~\equiv~ (F,G) - (F,\Theta^{a})E_{ab}(\Theta^{b},G)~,
\label{diracbracket}
\eeq
or in coordinates,
\beq
E^{AB}_{\DD} ~\equiv~ E^{AB} 
-(\Gamma^{A},\Theta^{a})E_{ab}(\Theta^{b},\Gamma^{B})~.
\label{diracbracketab}
\eeq
The Dirac antibracket satisfies a strong Jacobi identity
\beq
\sum_{F,G,H~{\rm cycl.}}(-1)^{(\eps_{F}+1)(\eps_{H}+1)}
((F,G)_{\D}^{},H)_{\D}^{}~=~0~.
\label{jaciddirac}
\eeq
The adjective ``strong'' stresses the fact that the Jacobi identity holds
off-shell \wrt the second-class constraints \mb{\Theta^{a}}, \ie everywhere in
the phase space \mb{M}. There is a canonical Dirac two-form given by
\beq
E^{\D}~\equiv~E - \Hf d\Theta^{a} E_{ab}\wedge d\Theta^{b}~,
\eeq
or in local coordinates
\beq
E_{AB}^{\DD} ~\equiv~ E^{}_{AB}
-(\lpa{A}\Theta^{a})E_{ab}(\Theta^{b}\rpa{B})~.
\eeq
The two-form field \mb{E_{AB}^{\DD}} is compatible with the Dirac bracket, \ie
it satisfies the property \e{tripleeee}, but it is {\em not} necessarily
closed. Local coordinates \mb{\Gamma^{A}=\{\gamma^{A};\Theta^{a}\}}, where the
second-class constraints \mb{\Theta^{a}} are part of the coordinates, are
called {\em unitarizing} coordinates. In the physics terminology, the 
second-class constraints \mb{\Theta^{a}} represent unphysical degrees of
freedom, which can be eliminated from the system, \ie put to zero, to reveal 
a reduced submanifold \mb{\tilde{M}}, whose coordinates \mb{\gamma^{A}} 
constitute the true physical degrees of freedom. {\em Notation}: We use capital
roman letters $A$, $B$, $C$, $\ldots$ from the beginning of the alphabet as 
upper index for both the full and the reduced variables \mb{\Gamma^{A}} and 
\mb{\gamma^{A}}, respectively. A tilde ``\mb{\sim}'' over an object will 
denote the corresponding reduced object.

\noi
Unitarizing coordinates \mb{\Gamma^{A}=\{\gamma^{A};\Theta^{a}\}}, where
the second-class variables \mb{\Theta^{a}} and the physical variables
\mb{\gamma^{A}} are perpendicular to each other in the antibracket sense
\beq
(\gamma^{A},\Theta^{a})~=~0~,
\eeq
are called {\em transversal} coordinates. One may prove that transversal 
coordinate systems exist locally, although one might have to reparametrize the
\mb{\Theta^{a}} constraints in order to get to them,
\cf Subsection~\ref{secreparam} below.

\subsection{The Dirac Operators \mb{\DeltaED} and \mb{\DeltarhoED}}
\label{secdiracop}

\noi
The next step is to build Khudaverdian's BV operator \mb{\DeltaED} for the
degenerate Dirac antibracket structure \e{diracbracket}, and, if a density
\mb{\rho} is available, the odd Laplacian \mb{\DeltarhoED}. In other words,
one should substitute \mb{E\to\ED} everywhere in the previous
Section~\ref{secantipoissongeom}. Some facts about the \mb{\DeltaED} operator
are immediately clear. {}First of all, it is covariant under general coordinate
transformations, \cf Subsection~\ref{secarbi}. {}Furthermore, it is strongly
nilpotent 
\beq
\Delta_{\ED}^{2}~=~0~,\label{diracdeltaenilp}
\eeq
due to the strong Jacobi identity \e{jaciddirac} and
Theorem~\ref{theoremnilpjac}. The following Proposition~\ref{propositionnurhod}
expresses the Dirac odd scalar \mb{\nurhoED} in terms of the non-degenerate
antisymplectic structure and the second-class constraints \mb{\Theta^{a}}.

\begin{proposition}
The {\bf Dirac odd scalar} \mb{\nurhoED} is given by
\beq
\nurhoED~=~\nurho-\frac{\nu^{(6)}_{\rho,\D}}{2}-\frac{\nu^{(7)}_{\rho,\D}}{2}
-\frac{\nu^{(8)}_{\D}}{8}+\frac{\nu^{(9)}_{\D}}{24}~,\label{nurhod}
\eeq
where \mb{\nurho\equiv\nurhoE} is the odd scalar for the non-degenerate
antisymplectic structure \mb{E}, and
\bea
\nu^{(6)}_{\rho,\D}
&\equiv&(\Deltarho\Theta^{a})E_{ab}(\Deltarho\Theta^{b})(-1)^{\eps_{b}}~,
\label{nurho6d} \\
\nu^{(7)}_{\rho,\D}
&\equiv&(-1)^{\eps_{a}+\eps_{b}}(\Theta^{a},E_{ab}(\Deltarho\Theta^{b}))
~=~(\Theta^{a},(\Deltarho\Theta^{b})E_{ba})~,\label{nurho7d} \\
\nu^{(8)}_{\D}&\equiv&(-1)^{\eps_{b}}(\Theta^{a},(\Theta^{b},E_{ba}))~,
\label{nu8d} \\
\nu^{(9)}_{\D}&\equiv&(-1)^{(\eps_{a}+1)(\eps_{d}+1)}
(\Theta^{d},E_{ab})E^{bc}(E_{cd},\Theta^{a}) \cr
&=&-(-1)^{\eps_{b}}(\Theta^{a},E^{bc})E_{cd}(\Theta^{d},E_{ba})~.\label{nu9d}
\eea
\label{propositionnurhod}
\end{proposition}

\noi
{\sc Proof of Proposition~\ref{propositionnurhod}}:~~Since both sides of
\eq{nurhod} are scalars under general coordinate transformations, it is
sufficient to work in Darboux coordinates for the non-degenerate \mb{E^{AB}}
structure. By straightforward calculation, one gets
\bea
\nu^{(0)}_{\rho,\D}&=&\nu_{\rho}^{(0)}
-(\Deltaone\Theta^{a})E_{ab}(\Theta^{b},\ln\sqrt{\rho})
-\frac{(-1)^{\eps_{a}}}{2\sqrt{\rho}}(\Theta^{a},E_{ab}(\Theta^b,\sqrt{\rho}))
~,\label{nurho0d} \\
\nu^{(1)}_{\D}&=&
-4(\Deltaone\Theta^{a})E_{ab}(\Deltaone\Theta^{b})(-1)^{\eps_{b}}
-4(-1)^{\eps_{a}+\eps_{b}} (\Theta^{a},E_{ab}(\Deltaone\Theta^{b}))\cr
&&-\nu^{(8)}_{\D}
-(-1)^{\eps_{a}}(\lpa{A}\Theta^{a},~E_{ab})(\Theta^{b},\Gamma^{A})
~,\label{nu1d} \\
\nu^{(2)}_{\D}&=&(-1)^{\eps_{b}\eps_{c}}E_{ca}
(\Theta^{a},~\Theta^{b}\rpa{B})E^{BC}_{\DD}
(\lpa{C}\Theta^{c},~\Theta^{d})E_{db} \cr
&=&-(-1)^{\eps_{a}}(\Theta^{b},~\Theta^{a}\rpa{A})
(\Gamma^{A},E_{ab})_{\D}^{}
~=~-(-1)^{\eps_{a}}(\Theta^{b},~\Theta^{a}\rpa{A})
(\Gamma^{A},E_{ab})-\frac{\nu^{(9)}_{\D}}{3}~,\label{nu2d} \\
\nu^{(3)}_{\D}&=&0~,\label{nu3d} \\
\nu^{(4)}_{\D}&=&0~,\label{nu4d} \\
\nu^{(5)}_{\D}&=&0~.\label{nu5d} 
\eea
The pertinent linear combination \e{nurho} of eqs.\ \e{nurho0d}--\e{nu5d} 
yields the \eq{nurhod}.
\proofbox

\subsection{Annihilation Relations}
\label{secannrel}

\noi
The fact that the \mb{\Theta^{a}} constraints are null-directions for the
Dirac construction is reflected slightly differently in 1) the Dirac
antibracket \mb{(\cdot,\cdot)_{\D}^{}}, 2) the Dirac odd Laplacian
\mb{\DeltarhoED}, and 3) the \mb{\DeltaED} operator. Explicitly, for a scalar
function \mb{F}, a density \mb{\rho} and a semidensity \mb{\sigma}, one has 
\bea
(F,\Theta^{a})_{\D}^{}&=&0~,\label{thetakill1} \\
(\DeltarhoED\Theta^{a})&=&
\frac{(-1)^{\eps_{A}}}{2\rho}
\lpa{A}\rho(\Gamma^{A},\Theta^{a})_{\D}^{}~=~0~,
\label{thetakill2} \\
{}[\stackrel{\rightarrow}{\Delta}_{\ED}^{},\Theta^{a}]\sigma
&=&[\stackrel{\rightarrow}{\Delta}_{1,\ED}^{},\Theta^{a}]\sigma
~=~(\Delta_{1,\ED}^{}\Theta^{a})\sigma
+(-1)^{\eps_{a}}(\Theta^{a},\sigma)_{\D}^{}~=~0~, \label{thetakill3}
\eea
respectively. Eqs.\ \e{thetakill1}--\e{thetakill3} generalize to
\bea
(F,f(\Theta))_{\D}^{}&=&0~,\label{thetakill1f} \\
(\DeltarhoED f(\Theta))&=&0~,
\label{thetakill2f} \\
{}[\stackrel{\rightarrow}{\Delta}_{\ED}^{},f(\Theta)]\sigma&=&0~, 
\label{thetakill3f}
\eea
for an arbitrary function \mb{f(\Theta)} of the constraints \mb{\Theta^{a}}.
(In other words: \mb{f} is here assumed not to depend on the physical variables
\mb{\gamma^{A}}.) Note however, that if \mb{\Theta^{a}} is not among the
defining set of constraints, but only a linear combination of those (\ie the
coefficients in the linear combination could involve the physical variables
\mb{\gamma^{A}}), the last equality in each of the above eqs.\
\e{thetakill1}--\e{thetakill3f} becomes weak, \ie there could be off-shell
contributions, \cf next Subsection~\ref{secreparam} and \Ref{bbd06}.

\subsection{Reparametrization of Second-Class Constraints}
\label{secreparam}

\noi
A general and tricky feature of the Dirac construction, is, that it
{\em changes} if one uses another defining set of second-class constraints
\beq
\Theta^{a}~~~~~\longrightarrow~~~~~~ 
\Theta^{\prime a}~=~\Lambda^{a}{}_{b}(\Gamma)~\Theta^{b}~.
\eeq
However, the dependence is so soft that physics, which lives on-shell, is not
affected \cite{bbd06}. We shall here clarify in exactly what sense the
\mb{\DeltaED} operator remains invariant on-shell under reparametrization of
the constraints.

\noi
To warm up, let us recall that the Dirac antibrackets \mb{(F,G)_{\D}^{}} and
\mb{(F,G)_{\D}^{\prime}}, defined using the primed and unprimed constraints
\mb{\Theta^{\prime a}} and \mb{\Theta^{a}}, respectively, are the same on-shell
\beq
(F,G)_{\D}^{\prime}~\approx~(F,G)_{\D}^{}~.
\label{diracantibrackettransf0a}
\eeq
Here the symbol ``\mb{\approx}'' is the Dirac weak equivalence symbol, which
denotes equivalence modulo terms of order \mb{{\cal O}(\Theta)}. More
generally,
\beq
{}F^{\prime}~\approx~F~~\wedge~~G^{\prime}~\approx~G~~~~~~~~\Rightarrow~~~~~~~~
(F^{\prime},G^{\prime})_{\D}^{\prime}~\approx~(F,G)_{\D}^{}~.
\label{diracantibrackettransf0b}
\eeq 
Hence the reduced bracket 
\beq
(\tilde{F},\tilde{G})_{\sim}^{}~\equiv~\left. (F,G)_{\D}^{}\right|_{\Theta=0}~,
\eeq 
is independent of both the choice of constraints \mb{\Theta^{a}} and 
the representatives \mb{F=F(\Gamma)}, \mb{G=G(\Gamma)} on \mb{M}. Here
\mb{\tilde{F}\equiv F|_{\Theta=0}=\tilde{F}(\gamma)} and
\mb{\tilde{G}\equiv G|_{\Theta=0}=\tilde{G}(\gamma)} are functions on the
physical submanifold \mb{\tilde{M}}.

\noi
On the other hand, to have a well-defined notion of reduced densities and
semidensities on the physical submanifold \mb{\tilde{M}}, it is necessary 
to let the densities and semidensities transform as
\beq
\rho^{\prime}~\approx~ \rho~\Lambda~,~~~~~~~~
\sigma^{\prime}~\approx~ \sigma~\sqrt{\Lambda}~,
\label{sigmadtransformationrule1}
\eeq
under reparametrization of the defining set of constraints
\mb{\Theta^{a}\to\Theta^{\prime a}=\Lambda^{a}{}_{b}~\Theta^{b}}.
Here 
\beq
\Lambda~\equiv~\sdet (\Lambda^{a}{}_{b}) 
\eeq
denotes the superdeterminant of the reparametrization matrix 
\mb{\Lambda^{a}{}_{b}=\Lambda^{a}{}_{b}(\Gamma)}. The reduction
\beq
\tilde{\rho}~\equiv~\left. \rho \right|_{\Theta=0}~,~~~~~~~~
\tilde{\sigma}~\equiv ~\left. \sigma \right|_{\Theta=0}~,
\label{sigmaredux}
\eeq
is then by definition performed in a unitarizing coordinate system
\mb{\Gamma^{A}=\{\gamma^{A};\Theta^{a}\}}, where it is implicitly understood
that the \mb{\Theta^{a}} coordinates coincide with the defining set of
constraints. Similar to the Dirac antibracket \mb{(\cdot,\cdot)_{\D}^{}}, we
imagine that the densities and semidensities refer to an internal defining 
set of \mb{\Theta^{a}} constraints. If one chooses another defining set of
constraints \mb{\Theta^{\prime a}}, and an accompanying unitarizing coordinate
system \mb{\Gamma^{\prime A}=\{\gamma^{\prime A};\Theta^{\prime a}\}}, the 
superdeterminant factor \mb{\Lambda} in the reparametrization rule
\e{sigmadtransformationrule1} is designed to cancel the Jacobian factor \mb{J}
from the coordinate transformation \e{coordtransf} on-shell, so that the
reduced definition \e{sigmaredux} stays the same.

\noi
Similarly, it is necessary that the \mb{\DeltaED} operator, which takes
semidensities to semidensities, transforms as
\beq
\Delta^{\prime}_{E_{\D}^{}}~\approx~\sqrt{\Lambda}~\DeltaED
\frac{1}{\sqrt{\Lambda}}
\eeq
as an operator identity. Stated more precisely, the odd scalar \mb{\nurhoED}
from Definition~\ref{defnurho} should be invariant on-shell
\beq
\nu_{\rho^{\prime}, E^{~\prime}_{\D}}^{}~\approx~\nurhoED
\label{nurhodtransfrule}
\eeq
under reparametrization of the constraints. This is the core issue at stake. 
To prove that it indeed holds, first note that it is enough to check the claim
\e{nurhodtransfrule} if the set of unprimed constraints \mb{\Theta^{a}} happens
to belong to a set of transversal coordinates
\mb{\Gamma^{A}=\{\gamma^{A};\Theta^{a}\}}. (If this is not the case, one can
always locally find a transversal coordinate system, and split the above
reparamerization into two successive reparamerizations that both involve the
transversal coordinates.) Transversal coordinates will simplify considerably
the ensuing calculations. In general, the on-shell change of \mb{\nurhoED}
depends on how the Dirac antibracket \mb{(\cdot,\cdot)_{\D}^{}} changes up to
the second order in \mb{\Theta^{a}}, \cf eq.\ (4.39) in \Ref{bbd06}.
Explicitly, one may show that the quantities \mb{\nu^{(0)}_{\rho,\D}},
\mb{\nu^{(1)}_{\D}}, \mb{\nu^{(2)}_{\D}}, \mb{\nu^{(3)}_{\D}},
\mb{\nu^{(4)}_{\D}} and  \mb{\nu^{(5)}_{\D}}, defined in eqs.\
\e{nu1}--\e{nu5} and \e{nurho0}, transform as
\bea
\nu^{\prime(0)}_{\rho,\D}&\equiv&\frac{1}{\sqrt{\rho^{\prime}}}
(\Delta_{1,E^{~\prime}_{\D}}^{}\sqrt{\rho^{\prime}})
~\approx~\frac{1}{\sqrt{\Lambda \rho}}
(\Delta_{\frac{1}{\Lambda},E_{\D}^{}}\sqrt{\Lambda\rho})
~=~\nu^{(0)}_{\rho,\D}
-\sqrt{\Lambda}(\Delta_{1,E_{\D}^{}}^{}\frac{1}{\sqrt{\Lambda}})~,
\label{nurho0dtransfrule} \\
\nu^{\prime(1)}_{\D}&\approx&\nu^{(1)}_{\D}
+8\sqrt{\Lambda}(\Delta_{1,E_{\D}^{}}^{}\frac{1}{\sqrt{\Lambda}})
-(-1)^{\eps_{b}}(\papal{\Theta^{\prime a}}\Theta^{b},~
\papal{\Theta^{b}}\Theta^{\prime a})_{\D}^{}~,\label{nu1dtransfrule}   \\
\nu^{\prime(2)}_{\D}&\approx&\nu^{(2)}_{\D}
-(-1)^{\eps_{b}}(\papal{\Theta^{\prime a}}\Theta^{b},~
\papal{\Theta^{b}}\Theta^{\prime a})_{\D}^{}~, \label{nu2dtransfrule}  \\
\nu^{\prime(3)}_{\D}&\approx&\nu^{(3)}_{\D}~, \label{nu3dtransfrule}  \\
\nu^{\prime(4)}_{\D}&\approx&\nu^{(4)}_{\D}~, \label{nu4dtransfrule}  \\
\nu^{\prime(5)}_{\D}&\approx&\nu^{(5)}_{\D}~. \label{nu5dtransfrule}  
\eea
The last equality in \eq{nurho0dtransfrule} is a non-trivial property of the 
odd Laplacian. It is now easy to see that the relevant linear combination 
\mb{\nurhoED} of \mb{\nu^{(0)}_{\rho,\D}}, \mb{\nu^{(1)}_{\D}},
\mb{\nu^{(2)}_{\D}}, \mb{\nu^{(3)}_{\D}}, \mb{\nu^{(4)}_{\D}} and
\mb{\nu^{(5)}_{\D}} is invariant on-shell.

\subsection{Nilpotency Condition for the odd Dirac Laplacian \mb{\DeltarhoED}}
\label{secnilpdeltarhodirac}

\noi
One of the surprising conclusions of \Ref{bbd06} was that one cannot maintain
a strong nilpotency of the Dirac odd Laplacian \mb{\DeltarhoED} under
reparametrization of the second-class constraints. This is consistent with
our new results. Using the terminology of last Subsection~\ref{secreparam},
one would say that the effect is caused by the off-shell variations of the odd 
scalar \mb{\nurhoED} and the Dirac antibracket \mb{(\cdot,\cdot)_{\D}^{}}, \cf 
the following calculation:
\beq
\Delta_{\rho^{\prime}, E^{~\prime}_{\D}}^{2}
~=~(\nu_{\rho^{\prime}, E^{~\prime}_{\D}}^{},~\cdot~)^{\prime}_{\D}
~\approx~(\nurhoED,~\cdot~)_{\D}^{}~=~\Delta_{\rho,\ED}^{2}~.
\eeq
Here use is made of eqs.\ \e{deltarhosquared},
\es{diracantibrackettransf0b}{nurhodtransfrule}. This should be compared to the
situation with the \mb{\DeltaED} operator where the strong nilpotency
\e{diracdeltaenilp} is manifest from the onset, regardless of which defining
set of \mb{\Theta^{a}} constraints is used.

\subsection{Dirac Partition Function}
\label{secpathintd}

\noi
As an application of the \mb{\DeltaED} operator, it is interesting to consider
the first-level Dirac partition function in the \mb{\lambda^{*}_{\alpha}\!=\!0}
gauge. A review of the first-level formalism can be found in \Ref{bbd06}. 
The partition function reads
\beq
{\cal Z}_{\D}^{}~=~\int\! [d\Gamma][d\lambda] \left.
\exp[\Ih (\WED+\XED)] \right|_{\lambda^{*}=0}\prod_{a}\delta(\Theta^{a})~,
\label{z1d}
\eeq
where \mb{\WED\!=\!\WED(\Gamma)} and
\mb{\XED\!=\!\XED(\Gamma;\lambda,\lambda^{*})} satisfy the Quantum Master
Equations
\bea
\DeltaED\exp[\Ih\WED]&=&0~,\label{qmewd} \\
((-1)^{\eps_{\alpha}}\papal{\lambda^{\alpha}}\papal{\lambda^{*}_{\alpha}}
+\DeltaED)\exp[\Ih\XED]&=&0~.\label{qmexd}
\eea
The formula \e{z1d} for the Dirac partition function \mb{{\cal Z}_{\D}^{}}
differs from the original formula \cite{bbd06,bbd97} by not depending on a
\mb{\rho}. Instead, the partition function \mb{{\cal Z}_{\D}^{}} is invariant
under general coordinate transformations and under reparametrization of the
\mb{\Theta^{a}} constraints because the Boltzmann semidensities
\mb{\exp[\Ih\WED]} and \mb{\exp[\Ih\XED]} transform according to
\es{coordtransf}{sigmadtransformationrule1}. Given an arbitrary density
\mb{\rho}, it is possible to introduce Boltzmann scalars 
\bea
\exp[\Ih\Wrho]&\equiv&\exp[\Ih\WED] / \sqrt{\rho}~, \label{expwrhod} \\
\exp[\Ih\Xrho]&\equiv&\exp[\Ih\XED] / \sqrt{\rho}~, \label{expxrhod}
\eea
which satisfy corresponding Modified Quantum Master Equations similar to 
\eq{mqme}.

\section{Conversion of Second-Class into First-Class}
\label{secconv}

\noi
Originally, the conversion of second-class constraints into first-class 
constraints was developed for even Poisson geometry
\cite{bf87,bff89,bt91,fl94}. Later it was adapted to anti-Poisson geometry in 
\Ref{bm97}, more precisely to the Dirac antibracket \mb{(\cdot,\cdot)_{\D}^{}}
and odd Laplacian \mb{\DeltarhoED}. In this Section~\ref{secconv} we develop
the anti-Poisson conversion method further and show that the Dirac
\mb{\DeltaED} operator from last Section~\ref{secsecondclass} can also be
derived via conversion.

\subsection{Extended Manifold \mb{\Mext}}
\label{secextmani}

\noi
As in Section~\ref{secsecondclass} the starting point is a general
non-degenerate antisymplectic manifold \mb{(M;E)} with a set of globally
defined second-class constraints \mb{\Theta^{a}=\Theta^{a}(\Gamma)}, which have
Grassmann parity \mb{\eps(\Theta^{a})\!=\!\eps_{a}}. We now consider a
cartesian product \mb{\Mext\equiv M\times V}, where \mb{(V;\om)} is a vector
space with a constant and non-degenerate antisymplectic metric, and such that
the dimension of \mb{V} is equal to the number of \mb{\Theta^{a}} constraints.
We will often identify \mb{M} with \mb{M\times \{0\}\subseteq\Mext}. The
extended manifold \mb{\Mext} has antisymplectic structure
\mb{\Eext\equiv E\oplus\om}.

\noi
Assume that points (\ie vectors) in the vector space \mb{V} are described by a 
set of coordinates \mb{\Phi_{a}} with Grassmann parity
\mb{\eps(\Phi_{a})\!=\!\eps_{a}\!+\!1}. 
{}For each set of local coordinates \mb{\Gamma^{A}} for the manifold \mb{M}, 
the extended manifold \mb{\Mext} will have local coordinates
\mb{\Gamma^{A}_{\ext}\!\equiv\!\{\Gamma^{A};\Phi_{a}\}}.
{\em Notation}: We use capital roman letters $A$, $B$, $C$, $\ldots$ from the
beginning of the alphabet as upper index for both the original and the extended
variables \mb{\Gamma^{A}} and \mb{\Gamma^{A}_{\ext}}, respectively. 
In detail, the extended antibracket \mb{(\cdot,\cdot)_{\ext}^{}} on \mb{\Mext}
reads
\bea
(\Gamma^{A},\Gamma^{B})_{\ext}^{}&\equiv&(\Gamma^{A},\Gamma^{B})~=~E^{AB}~, \\
(\Gamma^{A},\Phi_{a})_{\ext}^{}&\equiv&0~, \\
(\Phi_{a},\Phi_{b})_{\ext}^{}&\equiv&\om_{ab}~,~~~~~~~~~~~~ 
\eps(\om_{ab})~=~\eps_{a}\!+\!\eps_{b}\!+\!1~,
\eea
where, in particular, the antisymplectic matrix
\mb{\om_{ab}=-(-1)^{\eps_{a}\eps_{b}} \om_{ba}} does not depend on
\mb{\Gamma^{A}} nor on \mb{\Phi_{a}}. In other words, up to a constant matrix,
the \mb{\Phi_{a}} coordinates are global Darboux coordinates for the vector
space \mb{V}.

\subsection{First-Class Constraints \mb{T^{a}}}
\label{secfirstclass}

\noi
One next seeks Abelian first-class constraints
\mb{T^{a}\!=\!T^{a}(\Gamma;\Phi)} such that
\beq
 (T^{a}, T^{b})_{\ext}~=~0~,~~~~~~~~~~
\left. T^{a}\right|_{\Phi=0}~=~\Theta^{a}~.
\label{tcondition}
\eeq
Eq.\ \e{tcondition} is the defining relation for the conversion of second-class
constraints \mb{\Theta^{a}} into first-class constraint \mb{T^{a}}. The
first-class constraints \mb{T^{a}} are treated as power series expansions in
the \mb{\Phi_{a}} variables
\beq
T^{a}~=~\Theta^{a}
+\twotuborg{\X^{ab}_{L}\Phi_{b}}{\Phi_{b}\X^{ba}_{R}}
+\Hf\threetuborg{\Y^{abc}_{L}\Phi_{c}\Phi_{b}}{\Phi_{b}\Y^{bac}_{M}\Phi_{c}}
{\Phi_{b}\Phi_{c}\Y^{cba}_{R}}
+\frac{1}{6}\Z^{abcd}_{L}\Phi_{d}\Phi_{c}\Phi_{b}+ {\cal O}(\Phi^{4})~.
\label{texpan}
\eeq
The expressions \mb{\X^{ab}_{L}\Phi_{b} \equiv \Phi_{b}\X^{ba}_{R}} and
\mb{\Y^{abc}_{L}\Phi_{c}\Phi_{b}\equiv\Phi_{b}\Y^{bac}_{M}\Phi_{c}\equiv
\Phi_{b}\Phi_{c}\Y^{cba}_{R}} inside the curly brackets ``\mb{\{~\}}'' of
\eq{texpan} reflect various (equivalent) ways of ordering the \mb{\Phi^{a}} 
variables. The rules for shifting between the ordering prescriptions are
\bea
\X^{ab}_{L}&=&(-1)^{(\eps_{a}+1)(\eps_{b}+1)}\X^{ba}_{R}~, \\ 
(-1)^{(\eps_{a}+1)(\eps_{b}+1)}\Y^{bac}_{L}&=&\Y^{abc}_{M}
~=~(-1)^{(\eps_{b}+1)(\eps_{c}+1)}\Y^{acb}_{R}~.
\eea
One may show that a solution \mb{T^{a}} to the system \e{tcondition} exists, 
but that it is not unique. {}For instance, the condition on the 
\mb{\X^{ab}\!=\!\X^{ab}(\Gamma)} structure functions reads
\beq
E^{ad}~\equiv~(\Theta^{a},\Theta^{d})
~=~-\X^{ab}_{L}\om_{bc}\X^{cd}_{R}~.\label{teq0}
\eeq
The matrices \mb{\X^{ab}_{L}} and \mb{\X^{ab}_{R}} are necessarily invertible 
with inverse matrices \mb{\X_{ab}^{L}\!=\!(-1)^{\eps_{a}\eps_{b}}\X_{ab}^{R}},
since both \mb{E^{ab}\!\equiv\!(\Theta^{a},\Theta^{b}) } and 
\mb{\om_{ab}\!\equiv\!(\Phi_{a},\Phi_{b})_{\ext}^{}} in \eq{teq0} are
invertible. One may view \mb{\X^{ab}} as a Grassmann-odd vielbein between the
curved second-class matrix \mb{E^{ab}} and the flat metric \mb{\om_{ab}}.
At the next order in \mb{\Phi^{a}}, the condition on the 
\mb{\Y^{abc}\!=\!\Y^{abc}(\Gamma)} structure functions reads
\beq
(\Theta^{a},\X^{cb}_{R})+\X^{ad}_{L}\om_{de}\Y^{ecb}_{R}
+(\X^{ac}_{L},\Theta^{b})+\Y^{acd}_{L}\om_{de}\X^{eb}_{R}~=~0~,\label{teq1}
\eeq 
and so forth.

\subsection{Gauge Invariance}
\label{secgaugeinv}

\noi
The idea is now to view the first-class constraints \mb{T^{a}} as generators of
gauge symmetry and \mb{\Phi_{a}\!=\!0} as a particular gauge. We start by
defining gauge-invariant observables on the extended manifold \mb{\Mext}.

\begin{definition}
A scalar function  \mb{\bF\!=\!\bF(\Gamma;\Phi)}, a density
\mb{\brho\!=\!\brho(\Gamma;\Phi)} or a semidensity
\mb{\bsigma\!=\!\bsigma(\Gamma;\Phi)} on the extended manifold \mb{\Mext} is 
called a {\bf gauge-invariant extension} of a scalar function
\mb{F\!=\!F(\Gamma)}, a density \mb{\rho\!=\!\rho(\Gamma)} or a semidensity
\mb{\sigma\!=\!\sigma(\Gamma)} on the original manifold \mb{M}, if
the following conditions are satisfied
\bea
(\bF,T^{a})_{\ext}^{}~=&0~,~~~~~~~~~~\left. \bF\right|_{\Phi=0}&=~F~, 
\label{fcondition} \\
(\Deltabrho T^{a})~=&0~,~~~~~~~~~~
\left. \brho\right|_{\Phi=0}&=~\rho j~,\label{rhocondition} \\
{} [\stackrel{\rightarrow}{\Delta}_{\Eext}^{},T^{a}] \bsigma~=&0~,~~~~~~~~~~
\left. \bsigma\right|_{\Phi=0}&=~\sigma\sqrt{j}~,\label{sigmacondition}
\eea
respectively, where the \mb{j}-factor is defined in \eq{jdef} below.
\label{defgaugeinvext}
\end{definition}

\subsection{The \mb{j}-Factor}
\label{secjfactor}

\noi
The factor
\beq
 j~\equiv~\left. \bj \right|_{\Phi=0}~=~\sdet(\om_{ac}\X^{cb}_{R})
\label{jdef}
\eeq
is defined as the \mb{\Phi\!=\!0} restriction of the superdeterminant
\beq
 \bj~\equiv~\sdet(\Phi_{a},T^{b})_{\ext}^{}
~=~\int [d\bC][dC]~\exp\left[\Ih\bC^{a}(\Phi_{a},T^{b})_{\ext}^{}C_{b}\right]~,
~~~~~~~\eps(\bC^{a})~=~\eps_{a}\!+\!1~=~\eps(C_{a})~.
\label{bjdef}
\eeq
The \mb{j}-factor \e{jdef} is independent of the choice of \mb{\X^{ab}}
structure functions because of \eq{teq0}. It is a density for the vector space 
\mb{V} such that the corresponding volume form \mb{j[d\Phi]} on \mb{V} is
independent of the choice of coordinates \mb{\Phi_{a}}. In this way the
multiplication with \mb{j} in \eq{rhocondition} transforms a density \mb{\rho}
on the manifold \mb{M} into a density \mb{\rho j} for the extended manifold
\mb{\Mext\!\equiv\!M\times V}. The \mb{j}-factor is unique up to an overall
constant and can be physically explained as a Faddeev-Popov determinant, see
Subsection~\ref{secpathintext}.

\noi
Below we shall overwhelmingly justify the \mb{j}-factor in
Definition~\ref{defgaugeinvext}, in particular, through the Conversion
Theorem~\ref{theoremconversion}, but let us start by briefly mentioning a
curious implication. Consider what happens to the set of vielbein solutions
\mb{\X^{ab}_{L}} to \eq{teq0} under reparametrizations of the defining set of
second-class constraints
\mb{\Theta^{a}\to\Theta^{\prime a}=\Lambda^{a}{}_{b}~\Theta^{b}}. It is natural
to expect that there exists a bijective map
\mb{\X^{ab}_{L}\to\X^{\prime ab}_{L}} between the solutions such that
\beq 
\X^{\prime ac}_{L}~\approx~\Lambda^{a}{}_{b}\X^{bc}_{L}~,
\eeq
where ``\mb{\approx}'' denotes weak equivalence, \cf
Subsection~\ref{secreparam}. According to such map, the \mb{j}-factor would
transform as
\beq
 j^{\prime}~\approx~\Lambda~j~.\label{jtransformationrule}
\eeq
Recalling the transformation rule \e{sigmadtransformationrule1} for \mb{\rho},
this implies that the density \mb{\left.\brho\right|_{\Phi=0}\!=\!\rho j} on 
\mb{\Mext} changes with the {\em square} of \mb{\Lambda},
\beq
\left.\brho^{\prime}\right|_{\Phi=0}~
\approx~\Lambda^{2} \left.\brho\right|_{\Phi=0}~.
\label{brhozerotransformationrule}
\eeq
So while the \mb{j}-factor does indeed cancel the effect of changing the
\mb{\Phi_{a}} coordinates, it {\em doubles} the effect of changing the
second-class constraints \mb{\Theta^{a}}! Nevertheless, this doubling
phenomenon fits nicely with the rest of the conversion construction, \cf 
Subsection~\ref{secpathintext} below.

\subsection{Discussion of Gauge Invariance}
\label{secdiscgaugeinv}

\noi
Let us now justify the conditions \e{fcondition}--\e{sigmacondition}. The first
condition \e{fcondition} is simply the antisymplectic definition of gauge
invariance. As an example of condition \e{fcondition}, note that a first-class
constraint \mb{T^{a}=\bTheta^{a}} is a gauge-invariant extension of the
corresponding second-class constraint \mb{\Theta^{a}}. The other two conditions
\es{rhocondition}{sigmacondition} are a priori less obvious, but there are many
reasons to impose them: 

\begin{enumerate}

\item
The three conditions \e{fcondition}--\e{sigmacondition} are covariant \wrt
coordinate changes. 

\item
The conditions \e{fcondition}--\e{sigmacondition} are consistent with each
others, say, if one considers a density \mb{\rho^{\prime}=\rho F}, or a
semidensity \mb{\sigma=\sqrt{\rho}}. 

\item
The conditions \e{fcondition}--\e{sigmacondition} are natural counterparts of 
the annihilation properties \e{thetakill1}--\e{thetakill3}.

\item
One may show that there exist unique gauge-invariant extensions \mb{\bF},
\mb{\brho} and \mb{\bsigma} satisfying the condition \e{fcondition},
\e{rhocondition} and \e{sigmacondition}, respectively. 

\item
The extended antibracket \mb{(\cdot,\cdot)_{\ext}^{}}, the extended odd
Laplacian \mb{\Deltabrho\equiv\DeltabrhoEext}, the extended \mb{\DeltaEext}
operator and the extended odd scalar \mb{\nubrho\equiv\nubrhoEext} are
compatibly with the gauge-invariance conditions
\e{fcondition}--\e{sigmacondition}, \ie
\bea
(\bF\bG,T^{a})_{\ext}^{}&=&\bF(\bG,T^{a})_{\ext}^{}
+(-1)^{\eps_{F}\eps_{G}}\bG(\bF,T^{a})_{\ext}^{}~=~0~, \\
((\bF,\bG)_{\ext}^{},T^{a})_{\ext}^{}&=&(\bF,(\bG,T^{a})_{\ext}^{})_{\ext}^{}
+(-1)^{(\eps_{a}+1)(\eps_{G}+1)}
((\bF,T^{a})_{\ext}^{},\bG)_{\ext}^{}~=~0~, \\
(\Deltabrho\bF,T^{a})_{\ext}^{}
&=&\Deltabrho(\bF,T^{a})_{\ext}^{}
+(-1)^{\eps_{F}}(\bF,\Deltabrho T^{a})_{\ext}^{}~=~0~,\\
{}[\stackrel{\rightarrow}{\Delta}_{\Eext}^{},T^{a}](\DeltaEext\bsigma)
&=&(\DeltaEext T^{a}\DeltaEext\bsigma)
~=~(\DeltaEext [T^{a},\stackrel{\rightarrow}{\Delta}_{\Eext}^{}]\bsigma)
~=~0~, \\
(\nubrho,T^{a})_{\ext}^{}&=& (\Delta_{\brho}^{2}T^{a})~=~0~.
\eea
Here use is made of the ordinary Leibniz rule, the Jacobi identity \e{jacid}, 
the BV Leibniz rule \e{deltaleibnizantib}, the \eq{deltaenilp} and the 
\eq{deltarhosquared}, respectively.

\item 
The conditions \e{fcondition}--\e{sigmacondition} imply the Conversion
Theorem~\ref{theoremconversion} below. 

\end{enumerate}

\subsection{The Conversion Map}
\label{secconversionmap}

\noi
The gauge-invariant extension map 
\beq
  {\cal F}(M)~\ni~F~~\stackrel{\cong}{\longmapsto}~~
\bF~\in~{\cal F}(\Mext)_{\rm inv}^{}
\eeq
(which is \aka the {\em conversion map}) is an isomorphism of functions on
\mb{M} to gauge-invariant function on \mb{\Mext}, \cf point \mb{4} of the last
Subsection~\ref{secdiscgaugeinv}. The inverse conversion map is simply the
restriction to \mb{M},
\beq
  {\cal F}(\Mext)_{\rm inv}^{}~\ni~\bF~~\stackrel{\cong}{\longmapsto}~~
\left.\bF\right|_{\Phi=0}~\in~{\cal F}(M)~.
\eeq
The following Theorem~\ref{theoremconversion} is the heart of the conversion
method. It shows that the inverse conversion map transforms the extended model
into the Dirac construction.

\begin{theorem}
The restrictions to \mb{M} of the extended antibracket
\mb{(\cdot,\cdot)_{\ext}^{}}, the extended odd Laplacian
\mb{\Deltabrho\equiv\DeltabrhoEext}, the extended \mb{\DeltaEext} operator and
the extended odd scalar \mb{\nubrho\equiv\nubrhoEext} reproduce the 
corresponding Dirac constructions:
\bea
\left. (\bF,\bG)_{\ext}^{}\right|_{\Phi=0}&=&(F,G)_{\D}^{}~,\label{b2d1} \\ 
\left. (\DeltabrhoEext\bF)\right|_{\Phi=0}&=&(\DeltarhoED F)~,\label{b2d2} \\
\left. (\DeltaEext\bsigma)\right|_{\Phi=0}&=&\sqrt{j}(\DeltaED\sigma)~,
\label{b2d4} \\
\left. \nubrhoEext\right|_{\Phi=0}&=&\nurhoED~.\label{b2d3} 
\eea
\label{theoremconversion}
\end{theorem}

\noi
In principle, it is enough to prove \eq{b2d3}, since \eq{b2d3}
\mb{\Leftrightarrow} \eq{b2d4} \mb{\Rightarrow} \eq{b2d2} \mb{\Rightarrow}
\eq{b2d1}. Nevertheless, we shall give independent proofs of eqs.\ \e{b2d1},
\es{b2d2}{b2d3} in Appendix~\ref{appconvproof}. The following
Corollary~\ref{corollaryconversion} restates the conclusions of Conversion
Theorem~\ref{theoremconversion} using the forward conversion map.

\begin{corollary}
\bea
(F G)^{-}_{}&=& \bF\bG~, \label{d2b0} \\
((F,G)_{\D}^{})^{-}_{}&=&(\bF,\bG)_{\ext}^{}~, \label{d2b1} \\
(\DeltarhoED F)^{-}_{}&=&(\DeltabrhoEext\bF)~, \label{d2b2}\\
(\sqrt{j}\DeltaED\sigma)^{-}_{}&=&(\DeltaEext\bsigma)~, \label{d2b4} \\
(\nurhoED)^{-}_{}&=&\nubrhoEext~.\label{d2b3}
\eea
\label{corollaryconversion}
\end{corollary}

\noi
In particular, eqs.\ \es{b2d1}{d2b1} show that the conversion map is an 
isomorphism in the sense of anti-Poisson algebras between the Dirac 
anti-Poisson algebra \mb{({\cal F}(M); (\cdot,\cdot)_{\D}^{})} and the 
anti-Poisson algebra
\mb{({\cal F}(\Mext)_{\rm inv}^{}; (\cdot,\cdot)_{\ext}^{})} of gauge-invariant
functions on \mb{\Mext}.

\subsection{Extended Partition Function}
\label{secpathintext}

\noi
The first-level partition function in the \mb{\lambda^{*}_{\alpha}=0} gauge
reads 
\beq
{\cal Z}_{\ext}^{}~=~\int\! [d\Gamma_{\ext}^{}][d\lambda] \left.
\exp[\Ih (\WEext+\XEext)]\right|_{\lambda^{*}=0}
\frac{1}{\sdet(\chi_{a},T^{b})_{\ext}^{}}
\prod_{c}\delta(T^{c})\prod_{d}\delta(\chi_{d})~,
\label{z1ext}
\eeq
where \mb{\WEext\!=\!\WEext(\Gamma_{\ext})} and 
\mb{\XEext\!=\!\XEext(\Gamma_{\ext};\lambda,\lambda^{*})} 
satisfy the Quantum Master Equations
\bea
\DeltaEext\exp[\Ih\WEext]&=&0~,\label{qmewext} \\
((-1)^{\eps_{\alpha}}\papal{\lambda^{\alpha}}\papal{\lambda^{*}_{\alpha}}
+\DeltaEext)\exp[\Ih\XEext]&=&0~,\label{qmewxext}
\eea
and they are gauge invariant in the sense of condition \e{sigmacondition}:
\bea
{}[\stackrel{\rightarrow}{\Delta}_{\Eext}^{},T^{a}]
\exp[\Ih\WEext]~=&0~,~~~~~~~~~~
\left.\exp[\Ih\WEext] \right|_{\Phi=0}&=~\sqrt{j}\exp[\Ih \WED]~, \\
{}[\stackrel{\rightarrow}{\Delta}_{\Eext}^{},T^{a}]
\exp[\Ih\XEext]~=&0~,~~~~~~~~~~
\left.\exp[\Ih\XEext] \right|_{\Phi=0}&=~\sqrt{j}\exp[\Ih \XED]~. 
\eea
Here the Boltzmann semidensities \mb{\exp[\Ih\WED]} and \mb{\exp[\Ih\XED]}
satisfy the Quantum Master Equations \es{qmewd}{qmexd}, respectively. It is an
important fact that in the gauge \mb{\chi_{a}\!=\!\Phi_{a}}, the expression
\e{z1ext} for the extended partition function reduces to the Dirac partition
function \e{z1d}, \ie
\beq
{\cal Z}_{\ext}^{}~=~{\cal Z}_{\D}^{}~.
\eeq
Given a density \mb{\rho\!=\!\rho(\Gamma)} on \mb{M}, and a density
\mb{\brho\!=\!\brho(\Gamma_{\ext})} on \mb{\Mext} that satisfies
\eq{rhocondition}, it is possible to introduce Boltzmann scalars
\bea
\exp[\Ih\Wbrho]&\equiv&\exp[\Ih\WEext] / \sqrt{\brho}~, \\
\exp[\Ih\Xbrho]&\equiv&\exp[\Ih\XEext] / \sqrt{\brho}~,
\eea
which satisfy corresponding Modified Quantum Master Equations similar to 
\eq{mqme}. The Quantum Actions \mb{\Wbrho} and \mb{\Xbrho} defined this way 
are automatically gauge invariant
\bea
(\Wbrho,T^{a})_{\ext}^{}~=&0~,~~~~~~~~~~
\left. \Wbrho\right|_{\Phi=0}&=~\Wrho~, \\
(\Xbrho,T^{a})_{\ext}^{}~=&0~,~~~~~~~~~~
\left. \Xbrho\right|_{\Phi=0}&=~\Xrho~.
\eea
Here \mb{\Wrho} and \mb{\Xrho} are defined in \eqs{expwrhod}{expxrhod}, 
respectively.

\section{Conclusions}
\label{secconc}

\noi
We have shown for a general degenerate anti-Poisson manifold (under the
relatively mild assumption of a compatible two-form field) how to define in 
arbitrary coordinates the \mb{\DeltaE} operator, which takes semidensities to
semidensities, \cf Lemma~\ref{lemmaourdeltaesigma}. A large class of such
degenerate antibrackets are provided by the Dirac antibracket construction.
We have given a formula for the Dirac \mb{\DeltaED} operator, \cf
Proposition~\ref{propositionnurhod}, and shown in Subsection~\ref{secreparam}
that it is on-shell invariant under reparametrizations of the second-class
constraints. {}Finally, we showed that the Dirac \mb{\DeltaED} operator also
follows from the antisymplectic conversion scheme, \cf Conversion
Theorem~\ref{theoremconversion}.

\noi
Let us conclude with the following remark. It is often pointed out that
the antibracket \mb{(\cdot,\cdot)} is a descendant of the odd Laplacian
\mb{\Deltarho}. It measures the failure of the odd Laplacian \mb{\Deltarho}
to act as a linear derivation, \ie to satisfy the Leibniz rule. It can be
written as a double-commutator \cite{bbd96,bda96,b06cmp}
\beq
(F,G)~=~(-1)^{\eps_{F}}[[\stackrel{\rightarrow}{\Delta}_{\rho}^{},F],G]1~.
\label{abviadeltarho}
\eeq
In turn, the odd Laplacian \mb{\Deltarho} is a descendant of the \mb{\DeltaE} 
operator \cite{bbd06,b06}
\beq
(\Deltarho F)~=~
\frac{1}{\sqrt{\rho}}[\stackrel{\rightarrow}{\Delta}_{E}^{},F]\sqrt{\rho}~.
\label{deltarhoviadeltae}
\eeq
That is, one has schematically the following hierarchy: 
\beq
\begin{array}{rccccl}
&\DeltaE~{\rm operator}& \\ \\
&\Downarrow& \\ \\
&{\rm Odd~Laplacian}~\Deltarho&
&\Leftarrow&&{\rm Density}~\rho \\ \\
&\Downarrow& \\ \\
&{\rm Antibracket}~(\cdot,\cdot)&
\end{array}
\eeq 
Whereas the \mb{\DeltaE} operator is manifestly nilpotent, \cf 
Theorem~\ref{theoremnilpjac}, there is no fundamental reason to require the odd
Laplacian \mb{\Deltarho} to be nilpotent. (Of course, if \mb{\Deltarho} is not
nilpotent, the Boltzmann scalar \mb{\exp[\Ih\Wrho]} would in general have to
satisfy a Modified Quantum Master Equation with a non-trivial \mb{\nurho} term,
\cf \eq{mqme}. See also the recent preprint \cite{bb07}.) The Dirac odd
Laplacian \mb{\DeltarhoED} offers more evidence that nilpotency of the odd
Laplacian is not fundamental, at least not in its strong formulation, since in
this case the nilpotency can only be maintained weakly under reparametrizations
of the second-class constraints \mb{\Theta^{a}}, \cf \Ref{bbd06} and
Subsection~\ref{secnilpdeltarhodirac}.

\vspace{0.8cm}
\noindent
{\sc Acknowledgement:}~The Author thanks I.A.~Batalin and P.H.~Damgaard for
discussions. The Author also thanks the Erwin Schr\"odinger Institute for warm
hospitality. This work is supported by the Ministry of Education of the Czech
Republic under the project MSM 0021622409.

\appendix

\section{Proof of bi-Darboux Theorem~\ref{theorembidarboux}}
\label{appbidarboux}

\noi
If there exists an atlas of bi-Darboux coordinates, the two-form 
\mb{E=d\phi^{*}_{\alpha}\wedge d\phi^{\alpha}} is obviously closed. Now 
consider the other direction. Assume that the two-form \mb{E} is closed. Then
there locally exists a pre-antisymplectic one-form potential \mb{\vartheta}
such that
\beq
d\vartheta~=~E~.\label{d12form}
\eeq
Independently one knows that locally there exist Darboux coordinates
\mb{\Gamma^{A}\!=\!\left\{\phi^{\alpha};\phi^{*}_{\alpha};\Theta^{a}\right\}}.
Since the two-form \mb{E} is assumed to be compatible with the anti-Poisson
structure, it must be of the form \e{emn}. It is always possible to organize
the pre-antisymplectic one-form potential as
\beq
 \vartheta~\sim~\phi^{*}_{\alpha} d\phi^{\alpha}+\vartheta_{A} d\gamma^{A}
+\vartheta^{\prime}_{a} d\Theta^{a}~,\label{vartheta1}
\eeq
where \mb{\gamma^{A}\!=\!\left\{\phi^{\alpha};\phi^{*}_{\alpha}\right\}}
collectively denotes the fields and the antifields without the Casimirs.
The symbol ``\mb{\sim}'' denotes equality modulo exact terms, whose precise
expressions are irrelevant, since we are ultimately only interested in the
two-form \mb{E}. It follows from eqs.\ \e{emn}, \es{d12form}{vartheta1} that
\beq
  (\papal{\gamma^{A}}\vartheta_{B})
~=~(-1)^{\eps_{A}\eps_{B}}(A \leftrightarrow B)~,
\eeq
and hence there locally exists a fermionic function \mb{\Psi^{\prime}} such 
that
\beq
\vartheta_{A}~=~(\papal{\gamma^{A}}\Psi^{\prime})~.
\eeq
Defining 
\beq
\vartheta_{a}~\equiv~\vartheta^{\prime}_{a}
-(\papal{\Theta^{a}}\Psi^{\prime})~,
\eeq
the pre-antisymplectic one-form potential \e{vartheta1} reduces to 
\beq
\vartheta~\sim~\phi^{*}_{\alpha} d\phi^{\alpha}+\vartheta_{a} d\Theta^{a}~.
\label{vartheta2}
\eeq
We would like to show that the second term \mb{\vartheta_{a} d\Theta^{a}} in
\eq{vartheta2} vanishes under a suitable anticanonical transformation. Eqs.\
\es{d12form}{vartheta2} imply that the matrices \mb{M_{a\alpha}} and 
\mb{N^{\alpha}{}_{a}} in \eq{emn} are
\bea
-M_{a\alpha}&=&(\vartheta_{a}\papar{\phi^{\alpha}})
~=~(\vartheta_{a},\phi^{*}_{\alpha})~, \\
N^{\alpha}{}_{a}&=&(\papal{\phi^{*}_{\alpha}}\vartheta_{a})
~=~(\phi^{\alpha},\vartheta_{a})~,
\eea
and that the pre-antisymplectic potential components 
\mb{\vartheta_{a}\!=\!\vartheta_{a}(\Gamma)} satisfy a flatness condition:
\beq
{}F_{ab}~\equiv~(\papal{\Theta^{a}} \vartheta_{b})
-(-1)^{\eps_{a}\eps_{b}}(\papal{\Theta^{b}}\vartheta_{a})
+(\vartheta_{a},\vartheta_{b})~=~0~.
\label{varthetacondition}
\eeq
Put more illuminating, the condition \e{varthetacondition} implies that the
vector fields 
\beq
D_{a}~\equiv~\papal{\Theta^{a}}+\ad\vartheta_{a}
\eeq
commute
\beq
[D_{a},D_{b}]~=~\ad F_{ab}~=~0~.\label{flat}
\eeq
Here the adjoint action ``\mb{\ad}'' refers to the antibracket 
\mb{(\ad F)G\equiv(F,G)}, where \mb{F} and \mb{G} are functions. In other
words, \mb{\ad F} denotes the Hamiltonian vector field with Hamiltonian \mb{F}.
The vector fields \mb{D_{a}} are not 
Hamiltonian, although they do preserve the antibracket
\beq
D_{a}(F,G)~=~(D_{a}[F],G)+(-1)^{\eps_{a}(\eps_{F}+1)}(F,D_{a}[G])~,
\label{dapreserveab}
\eeq
\ie they are generators of anticanonical transformations that do not leave the
Casimirs invariant. It is an important fact that the \mb{D_{a}} are covariant
derivatives in the Casimir directions with a Lie algebra valued gauge potential
\mb{\ad\vartheta_{a}}. Here the Lie algebra is (a subalgebra of) the space
\mb{\Gamma(TM)} of vector fields, equipped with the commutator
\mb{[\cdot,\cdot]}, \ie the Lie bracket of vector fields. An infinitesimal
variation \mb{\delta\vartheta_{a}} of the pre-antisymplectic potential
components \mb{\vartheta_{a}} must satisfy 
\beq
D_{a}[\delta\vartheta_{b}]~=~(-1)^{\eps_{a}\eps_{b}}(a\leftrightarrow b)
\label{deltavarthetaclosed1}
\eeq
in order to respect the flatness condition \e{varthetacondition}. The last
\eq{deltavarthetaclosed1} implies in turn, that the only allowed infinitesimal
variations \mb{\delta\vartheta_{a}} are infinitesimal gauge transformations
\beq
\delta\vartheta_{a}~=~D_{a}[\delta\Psi]~, \label{gaugetransf}
\eeq
where \mb{\delta\Psi} is an infinitesimal fermionic gauge generator. The 
infinitesimal gauge transformation of the gauge potential \mb{\ad\vartheta_{a}}
is
\beq
\ad(\delta\vartheta_{a})~=~[D_{a},\ad(\delta\Psi)]~, \label{adgaugetransf}
\eeq
where use is made of \eq{dapreserveab}. Despite the appearance, the
\eq{adgaugetransf} is exactly the standard formula
\mb{\delta A_{\mu}=D_{\mu}\eps} for infinitesimal non-Abelian gauge
transformations. Any discrepancy is merely in notation, not in content. So one
can take advantage of well-known facts about non-Abelian gauge theory and \eg
Wilson-lines. In particular, the infinitesimal transformations
\es{gaugetransf}{adgaugetransf} generalize to finite gauge transformations. The
field strength (or curvature) is zero, \cf \eq{flat}, so the gauge potential
\mb{\ad\vartheta_{a}} is pure gauge. This means that there locally exists a 
gauge where the gauge potential vanishes identically,
\beq
\ad\vartheta_{a}~=~0~. \label{advarthetalignul}
\eeq
An infinitesimal gauge transformation \e{gaugetransf} may be implemented \wtho
a Hamiltonian vector field \mb{\ad(\delta\Psi)} with infinitesimal Hamiltonian
\mb{\delta\Psi}. Using the active picture, the Lie derivative of the 
pre-antisymplectic one-form potential \wrt the Hamiltonian vector field 
\mb{\ad(\delta\Psi)} is
\beq
{\cal L}_{\ad(\delta\Psi)}\vartheta~=~[i_{\ad(\delta\Psi)},d]\vartheta
~\sim~i_{\ad(\delta\Psi)}E
~=~(\delta\Psi\papar{\gamma^{A}})d\gamma^{a}
+(\delta\Psi,\vartheta_{a})d\Theta^{a}
~\sim~ -D_{a}[\delta\Psi]d\Theta^{a}~. \label{liederivvartheta}
\eeq
\ie by flowing along the Hamiltonian vector field \mb{\ad(\delta\Psi)}, one may
mimic (minus) the infinitesimal gauge transformation \e{gaugetransf}. More
generally, finite gauge transformations of \mb{\vartheta_{a}} are in one-to-one
correspondence with anticanonical transformations that leave the Casimirs
invariant. In particular, one may go to the trivial gauge \e{advarthetalignul}
where the \mb{\vartheta_{a}} themselves are Casimirs. The flatness condition
\e{varthetacondition} then reduces to
\beq
(\papal{\Theta^{a}}\vartheta_{b})
~=~(-1)^{\eps_{a}\eps_{b}}(a\leftrightarrow b)~,
\label{deltavarthetaclosed2}
\eeq
so there exists a fermionic Casimir function \mb{\Psi\!=\!\Psi(\Theta)} such 
that
\beq
\vartheta_{a}~=~(\papal{\Theta^{a}}\Psi)~,
\eeq 
and hence the second term in \eq{vartheta2} is just an exact term,
\beq
\vartheta_{a}d\Theta^{a}~=~d\Psi~\sim~0~.
\eeq 
This shows that there locally exists an anticanonical transformation that
leaves the Casimirs invariant, such that the two-form reduces to
\mb{E=d\phi^{*}_{\alpha}\wedge d\phi^{\alpha}}.
\proofbox

\section{Details from the Proof of Lemma~\ref{lemmanurho}}
\label{appnurho}

\subsection{Proof of \eq{dnu2}}
\label{appnu2}

\noi
The infinitesimal variation of \mb{\nu^{(2)}} in \eq{nu2} yields \mb{8} 
contributions to linear order in the variation \mb{\delta\Gamma^{A}=X^{A}}, 
which may be organized as \mb{2\times 4} terms
\beq
\delta\nu^{(2)}~=~2(-\delta\nu_{I}^{(2)}-\delta\nu_{II}^{(2)}
+\delta\nu_{III}^{(2)}+\delta\nu_{IV}^{(2)})~,
\eeq
due to a \mb{(A,B) \leftrightarrow (D,C)} symmetry in \eq{nu2}. They are
\bea
\delta\nu_{I}^{(2)}&\equiv&(-1)^{\eps_{A}\eps_{C}}(\lpa{D}
E^{AB})E_{BF}(X^{F}\rpa{C})(\lpa{A}E^{CD})~, \\
\delta\nu_{II}^{(2)}&\equiv&(-1)^{\eps_{A}\eps_{C}}(\lpa{D}
E^{AB})E_{BC}(\lpa{A}X^{F})(\lpa{F}E^{CD})~, \\
\delta\nu_{III}^{(2)}&\equiv&(-1)^{\eps_{A}\eps_{C}}(\lpa{D}
E^{AB})E_{BC}\lpa{A}\left((X^{C}\rpa{F})E^{FD}\right)
~=~\delta\nu_{I}^{(2)}+\delta\nu_{V}^{(2)}~, \\
\delta\nu_{IV}^{(2)}&\equiv&(-1)^{\eps_{A}\eps_{C}}(\lpa{D}
E^{AB})E_{BC}\lpa{A}\left(E^{CF} (\lpa{F}X^{D})\right)
~=~\delta\nu_{II}^{(2)}+\delta\nu_{VI}^{(2)}~, \\
\delta\nu_{V}^{(2)}&\equiv&(-1)^{\eps_{A}\eps_{C}}E^{FD}
(\lpa{D}E^{AB})E_{BC}(\lpa{A}X^{C}\rpa{F})
~=~-\delta\nu_{V}^{(2)}+\delta\nu_{VII}^{(2)}~,\label{dnu2v} \\
\delta\nu_{VI}^{(2)}&\equiv&(-1)^{\eps_{A}}(\lpa{D}E^{AB})
P_{B}{}^{C}(\lpa{C}\lpa{A}X^{D})~, \\
\delta\nu_{VII}^{(2)}&\equiv&(-1)^{\eps_{A}}P_{C}{}^{D}(\lpa{D}
E^{AB})(\lpa{B}\lpa{A}X^{C})~, 
\eea
where we have noted various relations among the contributions. The Jacobi
identity \e{ejacid} for \mb{E^{AB}} is used in the second equality of
\eq{dnu2v}. Altogether, the infinitesimal variation of \mb{\nu^{(2)}}
becomes
\beq 
\delta\nu^{(2)}~=~2\delta\nu_{VI}^{(2)}+\delta\nu_{VII}^{(2)}~,
\eeq
which is \eq{dnu2}.

\subsection{Proof of \eq{dnu3}}
\label{appnu3}

\noi
The infinitesimal variation of \mb{\nu^{(3)}} in \eq{nu3} yields \mb{6} 
contributions to linear order in the variation \mb{\delta\Gamma^{A}=X^{A}},
\beq
 \delta\nu^{(3)}
~=~\delta\nu_{I}^{(3)}+\delta\nu_{II}^{(3)}+\delta\nu_{III}^{(3)}
-\delta\nu_{IV}^{(3)}-\delta\nu_{V}^{(3)}-\delta\nu_{VI}^{(3)}~.
\eeq
They are
\bea
\delta\nu_{I}^{(3)}&\equiv&(-1)^{\eps_{B}}(\lpa{A}E_{BC})
(X^{C}\rpa{F})E^{FD}(\lpa{D}E^{BA})~, \\
\delta\nu_{II}^{(3)}&\equiv&(-1)^{\eps_{B}}(\lpa{A}E_{BC})
E^{CD}\lpa{D}\left((X^{B}\rpa{F})E^{FA}\right)
~=~\delta\nu_{VII}^{(3)}+\delta\nu_{VIII}^{(3)}~, \\
\delta\nu_{III}^{(3)}&\equiv&(-1)^{\eps_{B}}(\lpa{A}E_{BC})
E^{CD}\lpa{D}\left(E^{BF}(\lpa{F}X^{A})\right)~=~
\delta\nu_{IV}^{(3)}+\delta\nu_{IX}^{(3)}~, \\
\delta\nu_{IV}^{(3)}&\equiv&(-1)^{\eps_{B}}E^{CD}(\lpa{D}E^{BA})
(\lpa{A}X^{F})(\lpa{F}E_{BC})~, \\
\delta\nu_{V}^{(3)}&\equiv&(-1)^{\eps_{B}}E^{CD}(\lpa{D}E^{BA})
\lpa{A}\left((\lpa{B}X^{F})E_{FC}\right)
~=~\delta\nu_{VII}^{(3)}+\delta\nu_{X}^{(3)}~, \\
\delta\nu_{VI}^{(3)}&\equiv&(-1)^{\eps_{B}}E^{CD}(\lpa{D}E^{BA})
\lpa{A}\left(E_{BF}(X^{F}\rpa{C})\right)
~=~\delta\nu_{I}^{(3)}-\delta\nu_{XI}^{(3)}~, \\
\delta\nu_{VII}^{(3)}&\equiv&(-1)^{\eps_{A}(\eps_{B}+1)}
(\lpa{A}E_{BC})E^{CD}(\lpa{D}E^{AF})
(\lpa{F}X^{B}) \cr
&=&-(-1)^{\eps_{B}(\eps_{C}+1)}E^{CD}(\lpa{D}E^{BA})
(\lpa{A}E_{CF})(X^{F}\rpa{B})~, \\
\delta\nu_{VIII}^{(3)}&\equiv&(-1)^{\eps_{B}}(\lpa{A}E_{BC})
E^{CD}(\lpa{D}X^{B}\rpa{F})E^{FA}~, \\
\delta\nu_{IX}^{(3)}&\equiv&(-1)^{\eps_{A}(\eps_{B}+1)}(\lpa{A}
E_{BC})E^{CD}(\lpa{D}X^{A}\rpa{F})E^{FB}~, \\
\delta\nu_{X}^{(3)}&\equiv&(-1)^{\eps_{A}}P_{C}{}^{D}
(\lpa{D}E^{AB})(\lpa{B}\lpa{A}X^{C})~, \\
\delta\nu_{XI}^{(3)}&\equiv&(-1)^{\eps_{B}(\eps_{C}+1)}E^{CD}
(\lpa{D}E^{BA})(\lpa{A}\lpa{C}X^{F})E_{FB}
~=~-\delta\nu_{X}^{(3)}-\delta\nu_{XI}^{(3)}~, \label{dnu3xi}
\eea
where we have noted various relations among the contributions. The Jacobi
identity \e{ejacid} for \mb{E^{AB}} is used in the second equality of
\eq{dnu3xi}. Altogether, the infinitesimal variation of \mb{\nu^{(3)}}
becomes
\beq
\delta\nu^{(3)}~=~\delta\nu_{VIII}^{(3)}+\delta\nu_{IX}^{(3)}
-\frac{3}{2}\delta\nu_{X}^{(3)}~,
\eeq
which is \eq{dnu3}.

\subsection{Proof of \eq{dnu4}}
\label{appnu4}
\noi
The infinitesimal variation of \mb{\nu^{(4)}} in \eq{nu4} yields \mb{6}
contributions to linear order in the variation \mb{\delta\Gamma^{A}=X^{A}},
\beq
\delta\nu^{(4)}~=~-\delta\nu_{I}^{(4)}-\delta\nu_{II}^{(4)}
+\delta\nu_{III}^{(4)}+\delta\nu_{IV}^{(4)}+\delta\nu_{V}^{(4)}
-\delta\nu_{VI}^{(4)}~.
\eeq
They are
\bea
\delta\nu_{I}^{(4)}&\equiv&(-1)^{\eps_{B}}E^{CD}(\lpa{D}E^{BF})
P_{F}{}^{A}\lpa{A}\left((\lpa{B}X^{G})E_{GC}\right)
~=~-\delta\nu_{VII}^{(4)}+\delta\nu_{VIII}^{(4)}~, \\
\delta\nu_{II}^{(4)}&\equiv&(-1)^{\eps_{B}}E^{CD}(\lpa{D}E^{BF})
P_{F}{}^{A}\lpa{A}\left(E_{BG}(X^{G}\rpa{C})\right)
~=~\delta\nu_{III}^{(4)}-\delta\nu_{IX}^{(4)}~, \\
\delta\nu_{III}^{(4)}&\equiv&(-1)^{\eps_{B}}P_{F}{}^{A}(\lpa{A}
E_{BC})(X^{C}\rpa{G})E^{GD}(\lpa{D}E^{BF})~, \\
\delta\nu_{IV}^{(4)}&\equiv&(-1)^{\eps_{B}}P_{F}{}^{A}(\lpa{A}
E_{BC})E^{CD}\lpa{D}\left((X^{B}\rpa{G})E^{GF}\right)
~=~\delta\nu_{X}^{(4)}+\delta\nu_{XI}^{(4)}~,\\
\delta\nu_{V}^{(4)}&\equiv&(-1)^{\eps_{B}}P_{F}{}^{A}(\lpa{A}E_{BC})
E^{CD}\lpa{D}\left(E^{BG}(\lpa{G}X^{F})\right)
~=~\delta\nu_{VI}^{(4)}+\delta\nu_{XII}^{(4)}~,\\
\delta\nu_{VI}^{(4)}&\equiv&(-1)^{\eps_{B}}P_{G}{}^{A}(\lpa{A}
E_{BC})E^{CD}(\lpa{D}E^{BF})(\lpa{F}X^{G})~, \\
\delta\nu_{VII}^{(4)}&\equiv&(-1)^{\eps_{B}(\eps_{C}+1)}E^{CD}
(\lpa{D}E^{BF})P_{F}{}^{A}(\lpa{A}E_{CG})(X^{G}
\rpa{B})~, \\
\delta\nu_{VIII}^{(4)}&\equiv&(-1)^{\eps_{B}}(\lpa{A}
\lpa{B}X^{C})P_{C}{}^{D}(\lpa{D}E^{BF})P_{F}{}^{A}~, \\
\delta\nu_{IX}^{(4)}&\equiv&(-1)^{\eps_{B}(\eps_{C}+1)}E^{CD}
(\lpa{D}E^{BF})P_{F}{}^{A}(\lpa{A}\lpa{C}
X^{G})E_{GB}~=~-\delta\nu_{VIII}^{(4)}-\delta\nu_{XIII}^{(4)}~,\label{dnu4ix}\\
\delta\nu_{X}^{(4)}&\equiv&(-1)^{(\eps_{B}+1)\eps_{F}}P_{F}{}^{A}
(\lpa{A}E_{BC})E^{CD}(\lpa{D}E^{FG})
(\lpa{G}X^{B})~=~-\delta\nu_{VII}^{(4)}~, \\
\delta\nu_{XI}^{(4)}&\equiv&(-1)^{\eps_{B}}(\lpa{A}E_{BC})
E^{CD}(\lpa{D}X^{B}\rpa{F})E^{FA}~, \\
\delta\nu_{XII}^{(4)}&\equiv&(-1)^{(\eps_{B}+1)\eps_{F}}P_{F}{}^{A}
(\lpa{A}E_{BC})E^{CD}(\lpa{D}X^{F}\rpa{G})
E^{GB}~, \\
\delta\nu_{XIII}^{(4)}&\equiv&(-1)^{(\eps_{A}+1)\eps_{B}}E^{AD}
(\lpa{D}E^{BC})(\lpa{C}\lpa{A}X^{G})E_{GB}
~=~-\delta\nu_{XIII}^{(4)}-\delta\nu_{XIV}^{(4)}~, \label{dnu4xiii}  \\
\delta\nu_{XIV}^{(4)}&\equiv&(-1)^{\eps_{A}}P_{C}{}^{D}(\lpa{D}
E^{AB})(\lpa{B}\lpa{A}X^{C})~,
\eea
where we have noted various relations among the contributions. The Jacobi
identity \e{ejacid} for \mb{E^{AB}} is used in the second equality of
\eqs{dnu4ix}{dnu4xiii}. Altogether, the infinitesimal variation of 
\mb{\nu^{(4)}} 
becomes
\beq
\delta\nu^{(4)}~=~-\delta\nu_{VIII}^{(4)}+\delta\nu_{IX}^{(4)}
+\delta\nu_{XI}^{(4)}+\delta\nu_{XII}^{(4)}
~=~-2\delta\nu_{VIII}^{(4)}+\delta\nu_{XI}^{(4)}
+\delta\nu_{XII}^{(4)}+\Hf\delta\nu_{XIV}^{(4)}~,
\eeq
which is \eq{dnu3}.

\subsection{Proof of \eq{dnu5}}
\label{appnu5}

\noi
The infinitesimal variation of \mb{\nu^{(5)}} in \eq{nu5} yields \mb{8}
contributions to linear order in the variation \mb{\delta\Gamma^{A}=X^{A}},
\beq
\delta\nu^{(5)}~=~\delta\nu_{I}^{(5)}+\delta\nu_{II}^{(5)}
+\delta\nu_{III}^{(5)}-\delta\nu_{IV}^{(5)}-\delta\nu_{V}^{(5)}
-\delta\nu_{VI}^{(5)}+\delta\nu_{VII}^{(5)}-\delta\nu_{VIII}^{(5)}~.
\eeq
They are
\bea
\delta\nu_{I}^{(5)}&\equiv&(-1)^{(\eps_{A}+1)\eps_{B}}
(X^{A}\rpa{G}) E^{GD}(\lpa{D}E^{BC})(\lpa{C}
E_{AF})P^{F}{}_{B}~, \\
\delta\nu_{II}^{(5)}&\equiv&(-1)^{(\eps_{A}+1)\eps_{B}}(\lpa{C}
E_{AF})P^{F}{}_{B}E^{AD}\lpa{D}\left((X^{B}\rpa{G})E^{GC}
\right)~=~\delta\nu_{VIII}^{(5)}+\delta\nu_{IX}^{(5)}~, \\
\delta\nu_{III}^{(5)}&\equiv&(-1)^{(\eps_{A}+1)\eps_{B}}(\lpa{C}
E_{AF})P^{F}{}_{B}E^{AD}\lpa{D}\left(E^{BG}(\lpa{G}X^{C})
\right)~=~\delta\nu_{IV}^{(5)}-\delta\nu_{X}^{(5)}~, \\
\delta\nu_{IV}^{(5)}&\equiv&(-1)^{(\eps_{A}+1)\eps_{B}}E^{AD}
(\lpa{D}E^{BC})(\lpa{C}X^{G})(\lpa{G}E_{AF})
P^{F}{}_{B}~, \\
\delta\nu_{V}^{(5)}&\equiv&(-1)^{(\eps_{A}+1)\eps_{B}}P^{F}{}_{B} E^{AD}
(\lpa{D}E^{BC})\lpa{C}\left((\lpa{A}X^{G})
E_{GF}\right)~=~\delta\nu_{I}^{(5)}+\delta\nu_{XI}^{(5)}~, \\
\delta\nu_{VI}^{(5)}&\equiv&(-1)^{(\eps_{A}+1)\eps_{B}}P^{F}{}_{B} E^{AD}
(\lpa{D}E^{BC})\lpa{C}\left(E_{AG}(X^{G}
\rpa{F})\right)~=~\delta\nu_{VII}^{(5)}-\delta\nu_{XII}^{(5)}~, \\
\delta\nu_{VII}^{(5)}&\equiv&(-1)^{(\eps_{A}+1)\eps_{B}} E^{AD}
(\lpa{D}E^{BC})(\lpa{C}E_{AF})(X^{F}\rpa{G})
P^{G}{}_{B}~, \\
\delta\nu_{VIII}^{(5)}&\equiv&(-1)^{(\eps_{A}+1)\eps_{B}} E^{AD}
(\lpa{D}E^{BC})(\lpa{C}E_{AF})P^{F}{}_{G}(X^{G}
\rpa{B})~, \\
\delta\nu_{IX}^{(5)}&\equiv&(-1)^{(\eps_{A}+1)\eps_{B}}E^{AD}
(\lpa{D}X^{B}\rpa{G})E^{GC}(\lpa{C}E_{AF})
P^{F}{}_{B} \cr
&=&(-1)^{\eps_{B}\eps_{F}}P_{F}{}^{D}(\lpa{D}X^{B}
\rpa{G})E^{GC}(\lpa{C}E^{FA})E_{AB}
~=~\delta\nu_{XI}^{(5)}+\delta\nu_{XII}^{(5)}~,\label{dnu5ix} \\
\delta\nu_{X}^{(5)}&\equiv&(-1)^{(\eps_{A}+1)\eps_{C}}(\lpa{C}
E_{AB})E^{BF}(\lpa{F}X^{C}\rpa{G})E^{GA}~, \\
\delta\nu_{XI}^{(5)}&\equiv&(-1)^{(\eps_{A}+1)\eps_{B}}E^{AD}
(\lpa{D}E^{BC})(\lpa{C}\lpa{A}X^{G})E_{GB}~,\\
\delta\nu_{XII}^{(5)}&\equiv&(-1)^{\eps_{B}}(\lpa{A}
\lpa{B}X^{C})P_{C}{}^{D}(\lpa{D}E^{BF})P_{F}{}^{A}~, 
\eea
where we have noted various relations among the contributions. The Jacobi
identity \e{ejacid} for \mb{E^{AB}} is used in the third equality of
\eq{dnu5ix}. Altogether, the infinitesimal variation of \mb{\nu^{(5)}}
becomes
\beq
\delta\nu^{(5)}~=~\delta\nu_{IX}^{(5)}-\delta\nu_{X}^{(5)}
-\delta\nu_{XI}^{(5)}+\delta\nu_{XII}^{(5)}
~=~-\delta\nu_{X}^{(5)}+2\delta\nu_{XII}^{(5)}~,
\eeq
which is \eq{dnu5}.

\section{Proof of Conversion Theorem~\ref{theoremconversion}}
\label{appconvproof}

\subsection{The \mb{\bj} Superdeterminant}
\label{secbj}

\noi
Even-though it is only the \mb{j}-factor \e{jdef} and not the whole \mb{\bj} 
superdeterminant \e{bjdef} that enters the conversion map, it is nevertheless 
convenient to organize the discussion in terms of coefficient functions for 
(the logarithm of) the \mb{\bj} superdeterminant
\beq
\ln\bj~\equiv~
\bn~=~n+\twotuborg{n^{a}_{L}\Phi_{a}}{\Phi_{a}n^{a}_{R}}
+\Hf n^{ab}_{L}\Phi_{b}\Phi_{a}+{\cal O}(\Phi^{3})
~,~~~~~~~~~n~\equiv~\ln j~.\label{nexpan}
\eeq
By combining eqs.\ \e{texpan}, \es{bjdef}{nexpan}, one finds the first-order
coefficient functions \mb{n^{a}} to be
\beq
\begin{array}{rcccl}
n^{a}_{L}&=&(-1)^{\eps_{b}}\X_{bc}^{R}\Y^{cba}_{M}
&=&(-1)^{\eps_{b}+1}\X_{bc}^{L}\Y^{cba}_{L} \\
n^{a}_{R}&=&\Y^{abc}_{M} \X_{cb}^{L} (-1)^{\eps_{b}}
&=&\Y^{abc}_{R}\X_{cb}^{R}(-1)^{\eps_{b}+1}~.
\end{array} 
\label{n1eq}
\eeq
The second-order coefficient functions read
\beq
n^{cd}_{L}~=~(-1)^{\eps_{b}+1}\X_{ba}^{L}\Z^{abcd}_{L}
+(-1)^{(\eps_{a}+1)\eps_{c}}\X_{ab}^{R}\Y^{bce}_{R}\X_{ef}^{R}\Y^{fad}_{M}~.
\label{n2eq}
\eeq
In particular, the contracted second-order coefficient function is
\beq
(-1)^{\eps_{c}+1}n^{cd}_{L}\om_{dc}~=~\z^{(1)}-\y^{(2)}~,
\label{n2ceq}
\eeq
where we have introduced the following short-hand notation
\bea
\y^{(2)}&\equiv&(-1)^{\eps_{a}\eps_{f}}\X_{ab}^{R}\Y^{bfc}_{M}
\om_{cd}\Y^{dae}_{M}\X_{ef}^{L}~, \label{yyeq} \\
\z^{(1)}&\equiv&(-1)^{\eps_{b}+\eps_{c}}
\X_{ba}^{L}\Z^{abcd}_{L}\om_{dc}~.\label{zdef}
\eea
Since there is not a unique choice of the structure functions \mb{\X^{ab}}, 
\mb{\Y^{abc}}, \mb{\Z^{abcd}}, etc, one must apply the \mb{T^{a}} involution 
relation \e{tcondition} to eliminate their appearances. We have to wait until
Subsection~\ref{secynux} to completely eliminate all \mb{\Y^{abc}} appearances,
but we can do a first step in this direction. The quadratic \mb{\Y^{abc}}
dependence inside the odd \mb{\y^{(2)}} variable \e{yyeq} can be related to a
linear \mb{\Y^{abc}} dependence inside a new \mb{\y^{(1)}} variable as follows
\bea
0&\equi{\e{tcondition}}&
\Hf(-1)^{\eps_{a}+1}\Y^{bac}_{R}\X_{cd}^{R}\papal{\Phi_{d}} \left. 
\X_{ae}^{L} (T^{e},T^{f})_{\ext}^{}\X_{fb}^{R} \right|_{\Phi=0} \cr
&=&\Hf (-1)^{\eps_{c}\eps_{e}}\X_{cd}^{R}\papal{\Phi_{d}} \left. 
(T^{e},T^{f})_{\ext}^{}\right|_{\Phi=0}\X_{fb}^{R}\Y^{bca}_{M}\X_{ae}^{L} \cr
&=&(-1)^{\eps_{b}}\Y^{cba}_{R}(\X_{ab}^{R},\Theta^{d})\X_{dc}^{R}
+(-1)^{\eps_{a}\eps_{f}}\X_{ab}^{R}\Y^{bfc}_{M}
\om_{cd}\Y^{dae}_{M}\X_{ef}^{L}~=~\y^{(1)}+\y^{(2)}~,
\eea
where
\beq
\y^{(1)}~\equiv~
(-1)^{\eps_{b}}\Y^{cba}_{R}(\X_{ab}^{R},\Theta^{d})\X_{dc}^{R}~.
\label{yeq}
\eeq
The only way the \mb{\Z^{abcd}_{L}} structure functions enters the discussion
is through the odd \mb{\z^{(1)}} variable \e{zdef}. It can be eliminated using
the following equation
\bea
0&\equi{\e{tcondition}}&
\X_{dc}^{R}\papal{\Phi_{c}} (-1)^{\eps_{a}}\X_{ab}^{R}\papal{\Phi_{b}}
\left.(T^{a},T^{d})_{\ext}^{}\right|_{\Phi=0} \cr
&=&(-1)^{\eps_{a}}\X_{ab}^{R}\papal{\Phi_{b}}\left.(T^{a},T^{d})_{\ext}^{}
\papar{\Phi_{c}}\X_{cd}^{L}(-1)^{\eps_{d}}\right|_{\Phi=0} \cr
&=&(n,n)+n^{a}_{L}\om_{ab}n^{b}_{R}
+2(-1)^{\eps_{b}+\eps_{c}}\X_{ba}^{L}(\Y^{abc}_{L},\Theta^{d})\X_{dc}^{R}
+(-1)^{\eps_{a}\eps_{d}}\X_{ab}^{R}(\X^{bd}_{R},\X^{ac}_{L})\X_{cd}^{L} \cr
&&+2(-1)^{\eps_{b}+\eps_{c}}\X_{ba}^{L}\Z^{abcd}_{L}\om_{dc}
+(-1)^{\eps_{a}\eps_{f}}\X_{ab}^{R}\Y^{bfc}_{M}
\om_{cd}\Y^{dae}_{M}\X_{ef}^{L} \cr
&=&(n,n)+n^{a}_{L}\om_{ab}n^{b}_{R}+2(n^{a}_{R},\Theta^{b})\X_{ba}^{R}
+(-1)^{\eps_{b}}(\X^{ab}_{L},\X_{ba}^{L})+2\z^{(1)}+\y^{(1)}~.\label{zeq}
\eea

\subsection{Gauge Invariant Function \mb{\bF}}
\label{secbf}

\noi
The gauge-invariant extension \mb{\bF} is a power series expansion in the 
\mb{\Phi_{a}} variables, \eg,
\beq
\bF~=~F+\twotuborg{\Phi_{a}F^{a}_{R}}{F^{a}_{L}\Phi_{a}}
+\Hf \Phi_{a}\Phi_{b}F^{ba}_{R}+{\cal O}(\Phi^{3})~. \label{fexpan}
\eeq
The coefficient functions for \mb{\bF} are uniquely determined by gauge
invariance condition \e{fcondition}. The first-order coefficient functions read
\beq
\begin{array}{rcccl}
{}F^{a}_{R}&=&-\om^{ab}\X_{bc}^{L} (\Theta^{c},F)
&=&\X^{ab}_{R} E_{bc}(\Theta^{c},F) ~, \\
{}F^{a}_{L}&=&-(F,\Theta^{c})\X_{cb}^{R}\om^{ba}
&=&(F,\Theta^{c})E_{cb}\X^{ba}_{L}~,
\end{array}
\label{f1eq} 
\eeq
The contracted second-order coefficient function 
\mb{(-1)^{\eps_{a}+1}\om_{ab}F^{ba}_{R}} is determined by the following 
calculation
\bea
0&\equi{\e{fcondition}}&(-1)^{\eps_{a}}\X_{ab}^{R}
\papal{\Phi_{b}}\left. (T^{a},\bF)_{\ext}^{}\right|_{\Phi=0} \cr
&=&\X_{ba}^{L}(\Theta^{a},F^{b}_{R})(-1)^{\eps_{b}+1}+(n,F)
+n^{a}_{L}\om_{ab}F^{b}_{R}+(-1)^{\eps_{a}+1}\om_{ab}F^{ba}_{R}~.\label{f2ceq} 
\eea

\subsection{Gauge Invariant Density \mb{\brho}}
\label{secbrho}

\noi
The (logarithm of the) gauge-invariant density \mb{\brho} is a power series
expansion in the \mb{\Phi_{a}} variables, \eg,
\beq
\ln\sqrt{\brho}~\equiv~
\bell~=~\ell+\twotuborg{\ell^{a}_{L}\Phi_{a}}{\Phi_{a}\ell^{a}_{R}}
+\Hf \Phi_{a}\Phi_{b}\ell^{ba}_{R}+{\cal O}(\Phi^{3})
~,~~~~~~~~~\ell~\equiv~\ln\sqrt{\rho j}~.\label{ellexpan}
\eeq
The coefficient functions for \mb{\brho} are uniquely determined by the gauge
invariance condition \e{rhocondition}. The first-order coefficient functions
\mb{\ell^{a}} can be found from the following Lemma~\ref{lnlemma}.

\begin{lemma}
\bea
\Hf n^{c}_{L}-\ell^{c}_{L}&=&
(\Deltarho\Theta^{a})\X_{ab}^{R}\om^{bc}
+\Hf(-1)^{\eps_{a}}(\Theta^{a},\X_{ab}^{R})\om^{bc} \cr
&=&\frac{(-1)^{\eps_{A}}}{2\rho}\lpa{A}
\rho(\Gamma^{A},\Theta^{a})\X_{ab}^{R}\om^{bc}~, \label{lneql} \\
\Hf n^{c}_{R}-\ell^{c}_{R}&=&
\om^{cb}\X_{ba}^{L} (\Deltarho\Theta^{a})(-1)^{\eps_{a}}
+\Hf\om^{cb}(\X_{ba}^{L},\Theta^{a})(-1)^{\eps_{a}} \cr
&=&\om^{cb}\X_{ba}^{L} (\Theta^{a},\Gamma^{A})\rho
\rpa{A}\frac{(-1)^{\eps_{A}}}{2\rho}~. \label{lneqr}
\eea
\label{lnlemma}
\end{lemma}

\noi
{\sc Proof of Lemma~\ref{lnlemma}}:~~Combine
\beq
0~\equi{\e{rhocondition}}~\left. (\Deltabrho T^{a}) \right|_{\Phi=0}
~=~(\Deltarhoj\Theta^{a})
+\Hf (-1)^{\eps_{b}+1}\om_{bc}\Y^{cba}_{R}
+\ell^{c}_{L}\om_{cb}\X^{ba}_{R}
\eeq
and
\bea
0&\equi{\e{tcondition}}&(-1)^{\eps_{b}}\X_{bc}^{R}
\papal{\Phi_{c}}\left.(T^{b},T^{a})_{\ext}^{}\right|_{\Phi=0} \cr
&=&(-1)^{\eps_{c}+1} \X_{cb}^{L}(\Theta^{b},\X^{ca}_{R})+(n,\Theta^{a})
+n^{c}_{L} \om_{cb}\X^{ba}_{R}+(-1)^{\eps_{b}+1}\om_{bc}\Y^{cba}_{R}~.
\eea
\proofbox

\noi
The contracted second-order coefficient function 
\mb{(-1)^{\eps_{a}+1}\om_{ab}\ell^{ba}_{R}} is determined by the following 
calculation
\bea
0&\equi{\e{rhocondition}}&\X_{ab}^{R}\papal{\Phi_{b}}\left.
(\Deltabrho T^{a})\right|_{\Phi=0} \cr
&=&(\Deltarhoj X^{ab}_{L}) \X_{ba}^{L}(-1)^{\eps_{a}}
+\Hf(-1)^{\eps_{b}+\eps_{c}}\X_{ba}^{L}\Z^{abcd}_{L}\om_{dc}\cr
&&+\X_{ba}^{L}(\Theta^{a},\ell^{b}_{L}) 
+n^{a}_{L}\om_{ab}\ell^{b}_{R}
+(-1)^{\eps_{a}+1}\om_{ab}\ell^{ba}_{R}~. \label{elleq2c}
\eea

\subsection{Assembling the Proof}
\label{secassemblingproof}

\noi
{\sc Proof of \eq{b2d1}}:
\beq
\left. (\bF,\bG)_{\ext}^{}\right|_{\Phi=0}
~=~(F,G)+F^{a}_{L} \om_{ab}G^{b}_{R}
~\equi{\e{f1eq}}~ (F,G)_{\D}^{}~.
\eeq
\proofbox

\noi
{\sc Proof of \eq{b2d2}}:
\bea
\left. (\DeltabrhoEext\bF)\right|_{\Phi=0}
&=&(\Deltarhoj F)+\Hf(-1)^{\eps_{a}+1} \om_{ab} F^{ba}_{R} 
+\ell^{a}_{L}\om_{ab}F^{b}_{R} \cr
&\equi{\e{f2ceq}}&(\Deltarhoj F)-\Hf(n,F)
+(\ell^{a}_{L}-\Hf n^{a}_{L})\om_{ab}F^{b}_{R}
-\Hf\X_{ba}^{L}(\Theta^{a},F^{b}_{R})(-1)^{\eps_{b}+1} \cr
&\equi{\e{lneql}}&(\Deltarho F)-(\Deltarho\Theta^{a})\X_{ab}^{R}F^{b}_{R}
-\Hf (-1)^{\eps_{a}}(\Theta^{a},\X_{ab}^{R}F^{b}_{R}) \cr
&=&(\Deltarho F)-(\Deltarho\Theta^{a})E_{ab}(\Theta^{b},F)
-\Hf (-1)^{\eps_{a}}(\Theta^{a},E_{ab}(\Theta^{b},F)) \cr
&=&(\DeltarhoED F)~.
\eea
\proofbox

\noi
{\sc Proof of \eq{b2d3}}:~~Using \eq{nurhof} it follows that 
\bea
\nurhoj-\nurho&\equi{\e{nurhof}}&\frac{1}{\sqrt{j}}(\Deltarho\sqrt{j}) 
~=~\Hf(\Deltarho n)+\frac{1}{8}(n,n) 
~=~\Hf(\Deltarhoj n) -\frac{1}{8}(n,n) \cr
&=&\Hf(\Deltarhoj X^{ab}_{L}) \X_{ba}^{L}(-1)^{\eps_{a}}
-\frac{1}{4}(-1)^{\eps_{b}}(\X^{ab}_{L},\X_{ba}^{L}) -\frac{1}{8}(n,n)~,
\label{nurhoj}
\eea
so that
\bea
\left. \nubrhoEext\right|_{\Phi=0}
&=&\nurhoj+\Hf(-1)^{\eps_{a}+1} \om_{ab} \ell^{ba}_{R}
+\Hf\ell^{a}_{L}\om_{ab}\ell^{b}_{R} \cr
&\equi{\e{nurhoj}}&\nurho-\frac{1}{8}(n,n)
+\Hf(\Deltarhoj X^{ab}_{L}) \X_{ba}^{L}(-1)^{\eps_{a}}
-\frac{1}{4}(-1)^{\eps_{b}}(\X^{ab}_{L},\X_{ba}^{L}) \cr
&&+\Hf(-1)^{\eps_{a}+1} \om_{ab} \ell^{ba}_{R}
+\Hf\ell^{a}_{L}\om_{ab}\ell^{b}_{R} \cr
&\equi{\e{elleq2c}}&\nurho-\frac{1}{8}(n,n)
-\Hf n^{a}_{L}\om_{ab}\ell^{b}_{R}
+\Hf\ell^{a}_{L}\om_{ab}\ell^{b}_{R} \cr
&&-\frac{1}{4}(-1)^{\eps_{b}}(\X^{ab}_{L},\X_{ba}^{L})
-\frac{\z^{(1)}}{4}-\Hf\X_{ba}^{L}(\Theta^{a},\ell^{b}_{L}) \cr
&\equi{\e{zeq}}&\nurho +\Hf(\Hf n^{a}_{L}
-\ell^{a}_{L})\om_{ab}(\Hf n^{b}_{R}-\ell^{b}_{R})
+\Hf(\Theta^{a},\Hf n^{b}_{L}-\ell^{b}_{L})\X_{ba}^{L} \cr
&&-\frac{1}{8}(-1)^{\eps_{b}}(\X^{ab}_{L},\X_{ba}^{L})+\frac{\y^{(1)}}{8}\cr
&\equi{\e{lneql}}&\nurho-\frac{\nu^{(6)}_{\rho,\D}}{2}
+\Hf(\Theta^{a},\X_{ab}^{R})\om^{bc}
(\Deltarho\Theta^{c})(-1)^{\eps_{c}}
+\frac{1}{8}(-1)^{\eps_{a}+\eps_{d}}
(\Theta^{a},\X_{ab}^{R})\om^{bc}(\X_{cd}^{L},\Theta^{d}) \cr
&&+\Hf(\Theta^{a},\Hf n^{b}_{L}-\ell^{b}_{L})\X_{ba}^{L} 
-\frac{1}{8}(-1)^{\eps_{b}}(\X^{ab}_{L},\X_{ba}^{L})+\frac{\y^{(1)}}{8}\cr
&=&\nurho-\frac{\nu^{(6)}_{\rho,\D}}{2}
-\frac{\nu^{(7)}_{\rho,\D}}{2}
+\frac{1}{8}(-1)^{\eps_{a}+\eps_{d}}
(\Theta^{a},\X_{ab}^{R})\om^{bc}(\X_{cd}^{L},\Theta^{d})\cr
&&+\frac{1}{4}(-1)^{\eps_{b}}
(\Theta^{a},(\Theta^{b},\X_{bc}^{R}))\om^{cd}\X_{da}^{L} 
-\frac{1}{8}(-1)^{\eps_{b}}(\X^{ab}_{L},\X_{ba}^{L})+\frac{\y^{(1)}}{8}\cr
&=&\nurhoED-\frac{\nu^{(9)}_{\D}}{24}-\frac{\x^{(1)}}{8}+\frac{\y^{(1)}}{8} 
~\equi{\e{ynux}}~\nurhoED~, \label{puttingalltogethernow}
\eea
where  the last equality in \eq{puttingalltogethernow} follows from
Lemma~\ref{ylemma} below, and the odd quantity \mb{\x^{(1)}} is defined in
\eq{x1eq}.
\proofbox

\subsection{Lemma~\ref{ylemma}}
\label{secynux}

\noi
It turns out that the most difficult part in the proof of \eq{b2d3} is to 
eliminate the \mb{\Y^{abc}} dependence from the odd \mb{\y^{(1)}} quantity 
\e{yeq}. Lemma~\ref{ylemma} gives a formula for \mb{\y^{(1)}} that are
manifestly independent of \mb{\Y^{abc}}. 

\begin{lemma}
\beq
\y^{(1)}~=~\frac{\nu^{(9)}_{\D}}{3}+\x^{(1)}~. \label{ynux}
\eeq
\label{ylemma}
\end{lemma}

\noi
{\sc Proof of Lemma~\ref{ylemma}}: ~~We first decompose the odd
\mb{\nu^{(9)}_{\D}} quantity \e{nu9d} as
\beq
\nu^{(9)}_{\D}~\equiv~(-1)^{(\eps_{a}+1)(\eps_{d}+1)}
(\Theta^{d},E_{ab})E^{bc}(E_{cd},\Theta^{a})
~=~-\x^{(1)}-2\x^{(2)}-\x^{(3)}~, \label{nu9dxeq}
\eeq
where
\bea
\x^{(1)}&\equiv&(-1)^{(\eps_{a}+1)(\eps_{d}+1)}
(\Theta^{d},\X_{ab}^{R})\om^{bc}(\X_{cd}^{L},\Theta^{a})~, \label{x1eq} \\
\x^{(2)}&\equiv&(-1)^{\eps_{b}(\eps_{d}+1)}
(\Theta^{d},\X^{bc}_{L})(\X_{cd}^{L},\Theta^{a})E_{ab}
~=~(-1)^{\eps_{b}(\eps_{d}+1)}
E_{ba}(\Theta^{a},\X_{dc}^{R})(\X^{cb}_{R},\Theta^{d})~,\label{x2eq} \\
\x^{(3)}&\equiv&(-1)^{\eps_{b}\eps_{e}}E_{ef}
(\Theta^{f},\X^{bc}_{L})\om_{cd}(\X^{de}_{R},\Theta^{a})E_{ab}~.\label{x3eq}
\eea
Secondly, we define
\bea
\x^{(4)}&\equiv&(-1)^{\eps_{c}}E_{ab}
(\Theta^{b},\X^{cd}_{L})(\X_{dc}^{L},\Theta^{a}) \cr
&=&-(-1)^{\eps_{c}}E_{ab}(\Theta^{b},\X^{cd}_{R})(\X_{dc}^{R},\Theta^{a})
~=~\x^{(3)}-\x^{(1)}~.\label{x4eq}
\eea
The third (=last) equality in \eq{x4eq} is a non-trivial assertion. To prove 
it, we define the following quantities:
\bea
\x^{(5)}&\equiv&(-1)^{\eps_{d}}(\Theta^{f},\X_{dc}^{R})
\X^{cb}_{R}E_{ba}(\Theta^{a},E^{de})E_{ef} \cr
&=&-(-1)^{\eps_{b}\eps_{e}}E_{ef}(\Theta^{f},\X^{bc}_{L})
\X_{cd}^{L}(E^{de},\Theta^{a})E_{ab}~=~\x^{(2)}+\x^{(3)}~,\label{x5eq} \\
\x^{(6)}&\equiv&(-1)^{\eps_{b}\eps_{e}}E_{ef}(\Theta^{f},\X^{bc}_{L})
\X_{cd}^{L}(\Theta^{d},E^{ea})E_{ab}~=~-\x^{(1)}-\x^{(2)}~,\label{x6eq} \\
\x^{(7)}&\equiv&(-1)^{\eps_{d}}(\Theta^{f},\X_{dc}^{R}) 
\X^{cb}_{R}E_{ba}(E^{ad},\Theta^{e})E_{ef}~=~-\x^{(4)}+\x^{(8)}~,\label{x7eq}\\
\x^{(8)}&\equiv&(-1)^{(\eps_{a}+1)(\eps_{d}+1)}
E^{ab}(\X_{bd}^{R},\Theta^{e})E_{ef}(\Theta^{f},\X_{ac}^{R})\om^{cd}
~=~0~,\label{x8eq}
\eea
where \eq{teq0} is used in the second equality of eqs.\ \e{x5eq},
\es{x6eq}{x7eq}. Remarkably the quantity \mb{\x^{(8)}} vanishes due to an
antisymmetry under the index permutation \mb{ace\leftrightarrow bdf}.
One may now check that the Jacobi identity 
\beq
 \sum_{{\rm cycl.}~a,b,c}(-1)^{(\eps_{a}+1)( \eps_{c}+1)}
(E^{ab},\Theta^{c}) ~=~0 \label{jacidtheta}
\eeq
yields \eq{x4eq}: 
\beq
0~=~\x^{(5)}+\x^{(6)}+\x^{(7)}~=~-\x^{(1)}+\x^{(3)}-\x^{(4)}~.\label{x134eq}
\eeq
Thirdly, we define 
\bea
\y^{(3)}&\equiv&
(-1)^{\eps_{c}}E_{ab}\Y^{bcd}_{L}\om_{de}\X^{ef}_{R}(\X_{fc}^{R},\Theta^{a})
~=~\y^{(1)}+\x^{(4)}+\x^{(2)}~, \label{y3eq} \\
\y^{(4)}&\equiv&
(-1)^{(\eps_{a}+1)\eps_{d}}\om^{ab}\X_{bc}^{L}\Y^{cde}_{L}\om_{ef}\X^{fg}_{R}
(\X_{ga}^{R},\Theta^{h})\X_{hd}^{R}
~=~\y^{(1)}-\x^{(1)}+\x^{(2)}~, \label{y4eq}
\eea
where \eq{teq1} is used in the second equality of \eqs{y3eq}{y4eq}. Note that
\mb{\x^{(1)}} to \mb{\x^{(8)}} are manifestly independent of the \mb{\Y^{abc}}
structure functions. We shall soon see that this is also the case for the
variables \mb{\y^{(1)}} to \mb{\y^{(4)}}. It turns out to be possible to
rewrite \mb{\y^{(3)}} as
\bea
\y^{(3)}&=&\Hf(-1)^{(\eps_{a}+1)(\eps_{c}+\eps_{f}+1)+\eps_{c}}
E_{ab}\Y^{bcd}_{L}\X_{de}^{L}(E^{ef},\Theta^{a})\X_{fc}^{R} \cr
&=&(-1)^{\eps_{a}\eps_{c}}
E_{ab}\Y^{bcd}_{L}\X_{de}^{L}(E^{ea},\Theta^{f})\X_{fc}^{R}
~=~-\y^{(1)}-\y^{(4)}~. \label{y143eq}
\eea
Here the Jacobi identity \e{jacidtheta} is used in the second equality of
\eq{y143eq}. Altogether, eqs.\ \e{y3eq}, \es{y4eq}{y143eq} yields
\beq
3\y^{(1)}~=~\x^{(1)}-2\x^{(2)}-\x^{(4)}~. \label{yxeq}
\eeq
Now Lemma~\ref{ylemma} follows by combining eqs.\ \e{nu9dxeq}, \es{x4eq}{yxeq}.
\proofbox


\begin{thebibliography}{999}

\bibitem{bv81} I.A.~Batalin and G.A.~Vilkovisky,
Phys.~Lett.\ {\bf 102B} (1981) 27.

\bibitem{bv83} I.A.~Batalin and G.A.~Vilkovisky,
Phys.~Rev.\ {\bf D28} (1983) 2567 [E: {\bf D30} (1984) 508].

\bibitem{bv84} I.A.~Batalin and G.A.~Vilkovisky, 
Nucl.~Phys.\ {\bf B234} (1984) 106.

\bibitem{b97} K.~Bering, arXiv:physics/9711010.

\bibitem{schwarz93} A.~Schwarz, Commun.~Math.~Phys.\ {\bf 155} (1993) 249.

\bibitem{bt93} I.A.~Batalin and I.V.~Tyutin, Int.~J.~Mod.~Phys.\ {\bf A8}
(1993) 2333.

\bibitem{bbd96} I.A.~Batalin, K.~Bering and P.H.~Damgaard, Phys.~Lett.\
{\bf B389} (1996) 673.

\bibitem{bbd06} I.A.~Batalin, K.~Bering and P.H.~Damgaard, Nucl.~Phys.\
{\bf B739} (2006) 389.

\bibitem{b06}
K.~Bering, J.~Math.~Phys.\ {\bf 47} (2006) 123513, arXiv:hep-th/0604117.

\bibitem{k99} O.M.~Khudaverdian, arXiv:math.DG/9909117.

\bibitem{kv02}
O.M.~Khudaverdian and Th.~Voronov, Lett.~Math.~Phys.\ {\bf 62} (2002) 127.

\bibitem{k02} O.M.~Khudaverdian, Contemp.~Math.\ {\bf 315} (2002) 199.

\bibitem{k04} O.M.~Khudaverdian, Commun.~Math.~Phys.\ {\bf 247} (2004) 353.

\bibitem{bbd97} I.A.~Batalin, K.~Bering and P.H.~Damgaard, Phys.~Lett.\
{\bf B408} (1997) 235.

\bibitem{bf87}
I.A.~Batalin and E.S.~Fradkin, Nucl.~Phys.\ {\bf B279} (1987) 514; 
Phys.~Lett.\ {\bf B180} (1986) 157; E: {\em ibid} {\bf B236} (1990) 528.

\bibitem{bff89} I.A.~Batalin, E.S.~Fradkin, and T.E.~Fradkina, 
Nucl.~Phys.\ {\bf B314} (1989) 158; E: {\em ibid} {\bf B323} (1989) 734; 
{\em ibid} {\bf B332} (1990) 723.

\bibitem{bt91}
I.A.~Batalin and I.V.~Tyutin, Int.~J.~Mod.~Phys.\ {\bf A6} (1991) 3599.

\bibitem{fl94} E.S.~Fradkin and V.Ya.~Linetskii, 
Nucl.~Phys.\ {\bf B431} (1994) 569; {\em ibid} {\bf B444} (1995) 577.

\bibitem{bm97} 
I.A.~Batalin and R.~Marnelius, Nucl.~Phys.\ {\bf B511} (1998) 495.

\bibitem{bda96}
K.~Bering, P.H.~Damgaard and J.~Alfaro, Nucl.~Phys.\ {\bf B478} (1996) 459.

\bibitem{b06cmp}
K.~Bering, Commun.~Math.~Phys.\ {\bf 274} (2007) 297; arXiv:hep-th/0603116.

\bibitem{bb07} I.A.~Batalin and K.~Bering, 
J.~Math.~Phys.\ {\bf 49} (2008) 033515, arXiv:0708.0400.

\end{thebibliography}
\end{document}